\documentclass[aps,pra,amsmath,amssymb,showpacs,twocolumn,nofootinbib]{revtex4-1}
\usepackage{graphicx,epsfig,epsf}
\usepackage{dcolumn}
\usepackage{bm}
\usepackage{mathbbol}
\usepackage{epstopdf}
\usepackage{mathtools}
\usepackage{comment}
\usepackage{graphicx}
\usepackage{dsfont}
\usepackage{amsfonts}
\usepackage{bbm}
\usepackage{xcolor}
\usepackage{stackrel}
\usepackage{multirow}
\makeatletter

\begin{document}

\newcommand{\bk}{\mathbf{k}}
\newcommand{\br}{\mathbf{r}}
\newcommand{\eps}{\boldsymbol{\varepsilon}}
\newcommand{\epsk}{\boldsymbol{\varepsilon}_\mathbf{k}}
\def\keps{\mathbf{k}\boldsymbol{\varepsilon}}

\newcommand{\matgamma}{\boldsymbol{\gamma}}
\newcommand{\matDelta}{\boldsymbol{\Delta}}

\newcommand{\tr}{{\rm Tr}}
\newcommand*{\etal}{\textit{et al.}}
\def\vec#1{\mathbf{#1}}
\def\ket#1{|#1\rangle}
\def\bra#1{\langle#1|}
\def\dm{\boldsymbol{\wp}}
\def\ldx{\eta_x}

\def\ftq#1{\mathcal{F}_{\mathbf{k}}\left[#1\right]}
\def\ftxiij#1{\mathcal{F}^{-1}_{\mathbf{r}}\left[#1\right]}
\def\ftxi#1{\mathcal{F}^{-1}_{\mathbf{r}}\left[#1\right]}

\def\ftqz#1{\mathcal{F}_{k_z}\left[#1\right]}
\def\ftxiz#1{\mathcal{F}^{-1}_{r_z}\left[#1\right]}

\newcommand{\sxb}{\sigma_x^{(j)}}
\newcommand{\sxa}{\sigma_x^{(i)}}
\newcommand{\syb}{\sigma_y^{(j)}}
\newcommand{\sya}{\sigma_y^{(i)}}
\newcommand{\szb}{\sigma_z^{(j)}}
\newcommand{\sza}{\sigma_z^{(i)}}

\newcommand{\spb}{\sigma_+^{(j)}}
\newcommand{\spa}{\sigma_+^{(i)}}
\newcommand{\smb}{\sigma_-^{(j)}}
\newcommand{\sma}{\sigma_-^{(i)}}
\newcommand{\smo}{\sigma_-}
\newcommand{\spl}{\sigma_+}

\newcommand{\cor}{\mathcal{C}_{ij}(t)}
\newcommand{\corcl}{\mathcal{C}_{ij}^{\mathrm{\,cl}}(t)}
\newcommand{\corem}{\mathcal{C}_{ij}^{\mathrm{em}}(t)}
\newcommand{\cork}{\mathcal{C}_{ij}(\mathbf{k},t)}
\newcommand{\corclk}{\mathcal{C}^\mathrm{\,cl}_{\mathbf{k}\eps} (t)}
\newcommand{\coremk}{\mathcal{C}^\mathrm{em}_{\mathbf{k}\eps} (t)}
\newcommand{\corkm}{\mathcal{C}_{ij}^\mathrm{ex}(\mathbf{k},t)}
\newcommand{\corkmz}{\mathcal{C}_{ij}^\mathrm{ex}(\mathbf{k})}
\newcommand{\cormz}{\mathcal{C}_{ij}^\mathrm{ex}}

\newcommand{\corkgs}{C_{ij}^\mathrm{\,gs}(\mathbf{k})}
\newcommand{\corkfs}{C_{ij}^\mathrm{\,Fock}(\mathbf{k})}
\newcommand{\corkth}{C_{ij}^\mathrm{\,th}(\mathbf{k})}

\newcommand{\corkzmz}{\mathcal{C}_{ij}^\mathrm{ex}(\mathbf{k}_0)}
\newcommand{\corkzzmz}{\mathcal{C}_{ij}^\mathrm{ex}(k_z,0)}
\newcommand{\corqmz}{\mathcal{C}_{ij}^\mathrm{ex}(\mathbf{q})}
\newcommand{\corqzmz}{\mathcal{C}_{ij}^\mathrm{ex}(q_z)}

\newcommand{\gij}{\gamma_{ij}}
\newcommand{\gijcl}{\gamma^{\mathrm{cl}}}
\newcommand{\gijclk}{\gamma^\mathrm{\,cl}(\mathbf{k}, \lambda)}
\newcommand{\gijemk}{\gamma^\mathrm{em}_{\mathbf{k}\eps} }
\newcommand{\giiclk}{\gamma^\mathrm{\,cl}(\mathbf{k}, \lambda)}
\newcommand{\giiemk}{\gamma^\mathrm{em}_{\mathbf{k}\eps} }
\newcommand{\gijclkz}{\gamma^\mathrm{\,cl}(\mathbf{k}_0, \lambda)}
\newcommand{\gijemkz}{\gamma^\mathrm{em}_{\mathbf{k}_0\eps} }
\newcommand{\giiclkz}{\gamma^\mathrm{\,cl}(\mathbf{k}_0, \lambda)}
\newcommand{\giiemkz}{\gamma^\mathrm{em}_{\mathbf{k}_0\eps} }
\newcommand{\tgijclkz}{\tilde{\gamma}^\mathrm{\,cl}(\mathbf{k}_0, \lambda)}

\newcommand{\gof}{\gamma}
\newcommand{\dof}{\Delta}

\newcommand{\Oij}{\Omega_{ij}}
\newcommand{\Oijclk}{\Omega^\mathrm{\,cl}(\mathbf{k}, \lambda)}
\newcommand{\Oijcl}{\Omega^\mathrm{\,cl}}
\newcommand{\Oijemk}{\Omega^\mathrm{em}_{\mathbf{k}\eps}}
\newcommand{\Oijem}{\Omega^\mathrm{em}}

\newcommand{\Dij}{\Delta_{ij}}
\newcommand{\Dijcl}{\Delta^{\mathrm{cl}}}
\newcommand{\Dijclk}{\Delta^\mathrm{\,cl}(\mathbf{k}, \lambda)}
\newcommand{\Dieml}{\Delta^{\mathrm{em}}}
\newcommand{\Dijemk}{\Delta^\mathrm{em}_{\mathbf{k}\eps} }

\newcommand{\Dijclxi}{\Delta^\mathrm{\,cl}(\xi_{ij})}
\newcommand{\Dijxi}{\Delta_{ij}(\boldsymbol{\xi}_{ij})}
\newcommand{\Dijxiz}{\Delta_{ij}(\xi_{z})}
\newcommand{\Dijclxizp}{\Delta^\mathrm{\,cl}(\xi_{z}')}

\title{Master equation for collective spontaneous emission with quantized atomic motion}

\date{\today}

\author{Fran\c{c}ois Damanet$^{1}$, Daniel Braun$^{2}$ and John Martin$^1$}

\affiliation{$^1$Institut de Physique Nucl\'eaire, Atomique et de
Spectroscopie, Universit\'e de Li\`ege, B\^at.\ B15, B - 4000
Li\`ege, Belgium}
\affiliation{$^2$Institut f\"ur theoritische Physik, Universit\"at T\"ubingen, 72076 T\"ubingen, Germany}

\begin{abstract}

We derive a markovian master equation for the internal dynamics of an ensemble of two-level atoms including all effects related to the quantization of their motion. Our equation provides a unifying picture of the consequences of recoil and indistinguishability of atoms beyond the Lamb-Dicke regime on both their dissipative and conservative dynamics, and applies equally well to distinguishable and indistinguishable atoms. We give general expressions for the decay rates and the dipole-dipole shifts for any motional states, and we find closed-form formulas for a number of relevant states (Gaussian states, Fock states and thermal states). In particular, we show that dipole-dipole interactions and cooperative photon emission can be modulated through the external state of motion.

\end{abstract}


\maketitle

\section{Introduction}

Spontaneous emission of light from initially excited atoms became one of the corner stones of our understanding of the interaction of light and matter, soon after the introduction of the ``photon''. It was introduced phenomenologically by Einstein \cite{Ein16} through his famous $A$-coefficient that gives the rate of spontaneous de-excitation of an excited atom. Later, spontaneous emission was understood through the theory of Wigner and Weisskopf \cite{Wei30} as the result of the perturbation of an atom through the vacuum-fluctuations of the electromagnetic field surrounding the atom. The infinitely large number of modes involved in the process leads to effectively irreversible behavior. Once this mechanism was understood, it became clear that the rate with which the excitation of an atom in a given state decays is not a natural constant for this atom, but can be influenced by its environment. By engineering the mode-structure of the electromagnetic environment of an atom, in particular through modifying the density of states of the field modes at the resonance frequency, spontaneous emission can be enhanced (in the case of  an increased density of states), or reduced (in the opposite case), as first found by Purcell in the context of nuclear resonance \cite{Pur46}. This important insight is now routinely used in photonic crystals, where an electromagnetic band-structure can be designed at will and used for creating e.g.~a band gap around the resonance frequency, resulting in largely increased lifetime of an excited atom, inverted spin, exciton, or plasmonic excitation \cite{Byk72,Yab87,Joh87}.  \\

Even earlier, Dicke studied spontaneous emission of several atoms in close vicinity of each other, and found that in such a case spontaneous emission becomes a cooperative effect in which the amplitudes of all atoms emitting simultaneously interfere. Depending on the initial collective internal state of the atoms, emission can be largely enhanced (superradiance), or reduced (subradiance) \cite{Dic54}. Superradiance developed to a large research field in its own right \cite{Bon71,Bon71b,Nar74,Gla76,Gro82,Dev96,Bra98,Bra98b,Che05,Akk08,Bra11,Wie11,Opp14,Wie15,Bha15}, culminating recently in matter-wave superradiance in cold atomic gases \cite{Ino99}. It was soon realized that dipole-dipole interactions between atoms can significantly alter these cooperative processes \cite{Cof78,Fri74, Fre86, Fre87, Fen13, Ric90, Fri72}, but can also be exploited for a variety of purposes, such as the (partial) trapping of light \cite{Bie12} or the implementation of quantum gates using the dipole blockade~\cite{Luk01}.

In this paper, we reveal yet a third mechanism how spontaneous emission can be influenced: Collective emission can be largely ``quantum programmed'' by engineering the external quantum state of motion of the atoms. To this end, we derive a master equation that fully takes into account the quantum nature of the atomic motion and, when relevant, the indistinguishability of atoms. This is essential when the atoms form a Bose-Einstein condensate or are loaded in an optical lattice. For example, when two fermionic atoms are placed in the same potential well and motional state, one in the internal excited state and the other in the ground state, the Pauli exclusion principle forbids the main decay channel, and leads to an increased lifetime of the atomic excited state~(see e.g.\ \cite{San11}). Moreover, it has been known for a long time that the coherence of radiation scattering off atoms in a solid (e.g.\ in X-ray or neutron scattering) can be influenced through the thermal motion of the atoms. This results in the Debye-Waller factor \cite{Deb13,Wal23} that describes the reduction of visibility of interference maxima as function of temperature.  But while in a solid one has in general little influence on the state of motion of the atoms (apart from controlling the temperature of the lattice), a whole new world has opened up in the physics of ion-traps and cold atoms. There, the external motional state can now be very well controlled and engineered, to the extent that quantum gates coupling internal states of the atoms originally relied heavily on the use of precise states of this external ``quantum bus'' \cite{Cir95}, even though this requirement could be relaxed later \cite{Mol99}. 

Thus, the quantum nature of the atomic motion appears to be an efficient way to influence the internal dynamics of atoms and its engineering has a wide range of potential applications~\cite{Lod04}. However, it turns out that most of the methods used to describe the internal dynamics of atoms including a quantum treatment of their motion are either restricted to the Lamb-Dicke regime \cite{Jav88,Bre95,Vog96,Lei03,Bra08,Pic10,Cer10} or do not account for both recoil and indistinguishability~\cite{Yin95,Dub96,Ber97,Mor99,Mcd07,Rog08}. Therefore, it appears worthwhile to develop a general theory of spontaneous emission of an ensemble of atoms valid for arbitrary quantum states of motion, which is the purpose of this paper. The master equation we derive constitutes a powerful tool to study the combined effects of the recoil and the indistinguishability of atoms on both their dissipative and conservative internal dynamics, even beyond the Lamb-Dicke regime. The dependence of the dipole-dipole interactions as well as the life-time under spontaneous emission on the motional state of the atoms might be observable in dense Rydberg gases, which are under intense current experimental and theoretical investigation \cite{Afr04,Rob04,Alt11,Pel14}. 

The paper is organized as follows. In Section II, we present our model. In section III, we derive a general master equation for the internal dynamics of atoms valid for arbitrary motional states. In section IV, we provide general expressions for the dipole-dipole shifts and decay rates which determine the conservative and dissipative part of the master equation, and discuss the effects of the indistinguishability of atoms on these quantities. In section V, we calculate explicitly the decay rates and the dipole-dipole shifts for particularly relevant motional states (Gaussian states, Fock states and thermal states), both for distinguishable and indistinguishable atoms. 

\section{Model and Hamiltonian}
We consider $N$ identical two-level atoms spontaneously emitting photons due to their interaction with the free electromagnetic field initially in vacuum, and treat their motion quantum-mechanically. 
In the point of view of Power-Zienau-Wolley (multipolar coupling scheme \cite{Buh12, Pow59, Coh87}), the Hamiltonian describing the composite system is
\begin{equation}\label{tHamiltonian}
    H= H_{A}+H_{F}+H_{AF},
\end{equation}
with $H_{A}$ the Hamiltonian of the atoms, $H_{F}$ the Hamiltonian of the free field, and $H_{AF}$ the interaction Hamiltonian responsible for emission/absorption of photons and field-mediated interactions between atoms. 

In Eq.~(\ref{tHamiltonian}), the atomic Hamiltonian $H_{A}=H_{A}^{\mathrm{ex}}+H_{A}^{\mathrm{in}}+ 
H_{A}^{\mathrm{self}}$ consists of an external, an internal and a self-interaction part, respectively given by
\begin{align}\label{tHAex}
&    H_{A}^{\mathrm{ex}} = \sum_{j=1}^N\left(\frac{\hat{\mathbf{p}}_{j}^2}{2M} + V(\hat{\mathbf{r}}_j) \right), \\
    \label{tHAin}
 &   H_{A}^{\mathrm{in}} =\frac{\hbar\omega_0}{2}\sum_{j=1}^N
    \sigma_z^{(j)}, \\
    \label{tHAself}
 &   H_{A}^{\mathrm{self}} =\frac{1}{2 \epsilon_0} \int |\hat{\mathbf{P}}\big(\mathbf{r}\big)|^2 \,d\mathbf{r} .
\end{align}
The external part $H_{A}^{\mathrm{ex}}$ corresponds to the kinetic and potential energy of the atoms, with $\hat{\mathbf{r}}_j$ and $\hat{\mathbf{p}}_j$ the center-of-mass position and momentum operators of atom $j$ ($j=1,\ldots,N$) of mass $M$ and $V(\mathbf{r})$ the external potential experienced by the atoms~\cite{footnote1}. We include the spin degree of freedom in the internal state and consider an external potential which does not depend on the spin. This form of $H_{A}^{\mathrm{ex}}$ is quite general and can account for a wide range of experimental settings. 
The internal part $H_{A}^{\mathrm{in}}$ of the atomic Hamiltonian corresponds to the internal energy of the atoms, with $\omega_0$ the atomic transition frequency and $\sigma_z^{(j)} =
|e_j\rangle \langle e_j|-|g_j \rangle \langle g_j|$ with $|g_j\rangle$ ($|e_j\rangle$) the lower (upper) level of atom $j$ of energy $-\hbar \omega_0/2$ ($\hbar \omega_0/2$). Finally, the self-interaction part $H_{A}^{\mathrm{self}}$ corresponds to the self-energy and contact interaction between atoms, with $\epsilon_0$ the permittivity of free space and $\hat{\mathbf{P}}\big(\mathbf{r}\big)$ the atomic polarization density, given in the dipole approximation by~\cite{Pow59}
\begin{equation}
\hat{\mathbf{P}}\big(\mathbf{r} \big) =  \sum_{j = 1}^N \mathbf{D}_j \, \delta(\mathbf{r}-\hat{\mathbf{r}}_j)
\end{equation}
where $\mathbf{D}_j=\mathbf{d}_j\,\sigma_-^{(j)}+\mathbf{d}_j^*\,\sigma_+^{(j)}$ is the dipole operator for atom $j$, with dipole matrix element $\mathbf{d}_j=\bra{g_j}\mathbf{D}_j\ket{e_j}$, $\sigma_-^{(j)}=|g_j\rangle\langle e_j|$, $\sigma_+^{(j)}=|e_j\rangle\langle g_j|$ and $\delta$ is the Dirac delta distribution.  We consider a polarized atomic sample in which all atoms share the same dipole moment, i.e.\ $\mathbf{d}_j=\mathbf{d}\;\forall\,j$. The dipole moment $\mathbf{d}$ can be decomposed in the spherical basis $\{\eps_0\equiv\mathbf{e}_z,\eps_\pm\equiv\mp (\mathbf{e}_x\pm i \mathbf{e}_y)/\sqrt{2}\}$ with $\{\mathbf{e}_x, \mathbf{e}_y,\mathbf{e}_z\}$ the Cartesian unit vectors and the $z$-axis taken as the quantization axis,
\begin{equation}\label{decompd}
\mathbf{d}=\sum_{q=0,\pm}d_q\,\eps_q.
\end{equation}
For a $\pi$ transition from the upper to the lower level, the only non-vanishing component in (\ref{decompd}) is $d_0$, whereas for a $\sigma^\pm$ transition, the only non-vanishing component is $d_\mp$.

In Eq.~(\ref{tHamiltonian}), the free field Hamiltonian $H_{F}$ reads
\begin{equation}  \label{tHF}
H_{F} = \sum_{\keps} \hbar\omega_k\, a^\dagger_{\keps} a_{\keps},
\end{equation}
with $\omega_k=ck$, $k=|\bk|$, $c$ the speed of light in vacuum and $a_{\keps}$
($a^\dagger_{\keps}$) the annihilation (creation) operator of a mode of the radiation field of wave vector $\mathbf{k}$ and polarization $\eps$. Note that in Eq.~(\ref{tHF}), we have dropped the zero-point energy of the radiation field, as it has no influence on the dynamics of the system.

In the dipole approximation (when the typical size of the atoms is much smaller than the wavelength of the emitted radiation) and the interaction picture with respect to $H_0\equiv H_{A}^{\mathrm{ex}}+H_{A}^{\mathrm{in}}+H_F$, the interaction Hamiltonian $H_{AF}(t)$ reads
\begin{equation}\label{tHAF}
    H_{AF}(t)=-\sum_{j=1}^N \mathbf{D}_j(t)\boldsymbol{\cdot}\mathbf{E}\big (\hat{\mathbf{r}}_j(t),t \big)
\end{equation}
with the electric field operator
\begin{equation}\label{E}
  \mathbf{E}(\mathbf{r},t)=i\sum_{\mathbf{k}\boldsymbol{\varepsilon}}{\cal E}_k\, \left(a_{\keps} \,\epsk \, e^{i(\mathbf{k}\boldsymbol{\cdot}\mathbf{r}-\omega_k t)} - \mathrm{h.c.}\right)
\end{equation}
where h.c.\ stands for Hermitian conjugate, ${\cal E}_k=\sqrt{\hbar
\omega_k/2\epsilon_0 L^3}$, $L^3$ is the electromagnetic mode quantization volume, $\epsk$
the normalized polarization vector, and
\begin{equation}\label{xt}
    \hat{\mathbf{r}}_j(t)= e^{i H_A^\mathrm{ex} t/\hbar} \,  \hat{\mathbf{r}}_j \,e^{- i H_A^\mathrm{ex} t/\hbar}.
\end{equation} 
Performing the Schmidt decomposition of the dipole interaction Hamiltonian~(\ref{tHAF}), we get \cite{Bre06}
\begin{equation}\label{HIgen}
   H_{AF}(t)=\sum_{j=1}^N\sum_{\omega=\pm\omega_0}e^{-i\omega t} A_{j}^{\mathrm{in}}(\omega)\otimes B_{j}(t),
\end{equation}
with the \emph{quantum jump operators}
\begin{equation} \label{Aj} 
\begin{aligned}
& A^{\mathrm{in}}_{j}(\omega_0)= \sigma_-^{(j)},\\
& A^{\mathrm{in}}_{j}(-\omega_0) = \sigma_+^{(j)},
\end{aligned}
\end{equation} 
and the \emph{bath operators}
\begin{equation} \label{Bj} 
B_j(t)=-\mathbf{d}\boldsymbol{\cdot}\mathbf{E}\big( \hat{\mathbf{r}}_{j}(t),t\big)
\end{equation}
defined for any atom $j=1,\ldots,N$.

\section{General master equation for the internal dynamics}\label{sec.ME}
\begin{figure}
\begin{center}
\includegraphics[width=8.5cm]{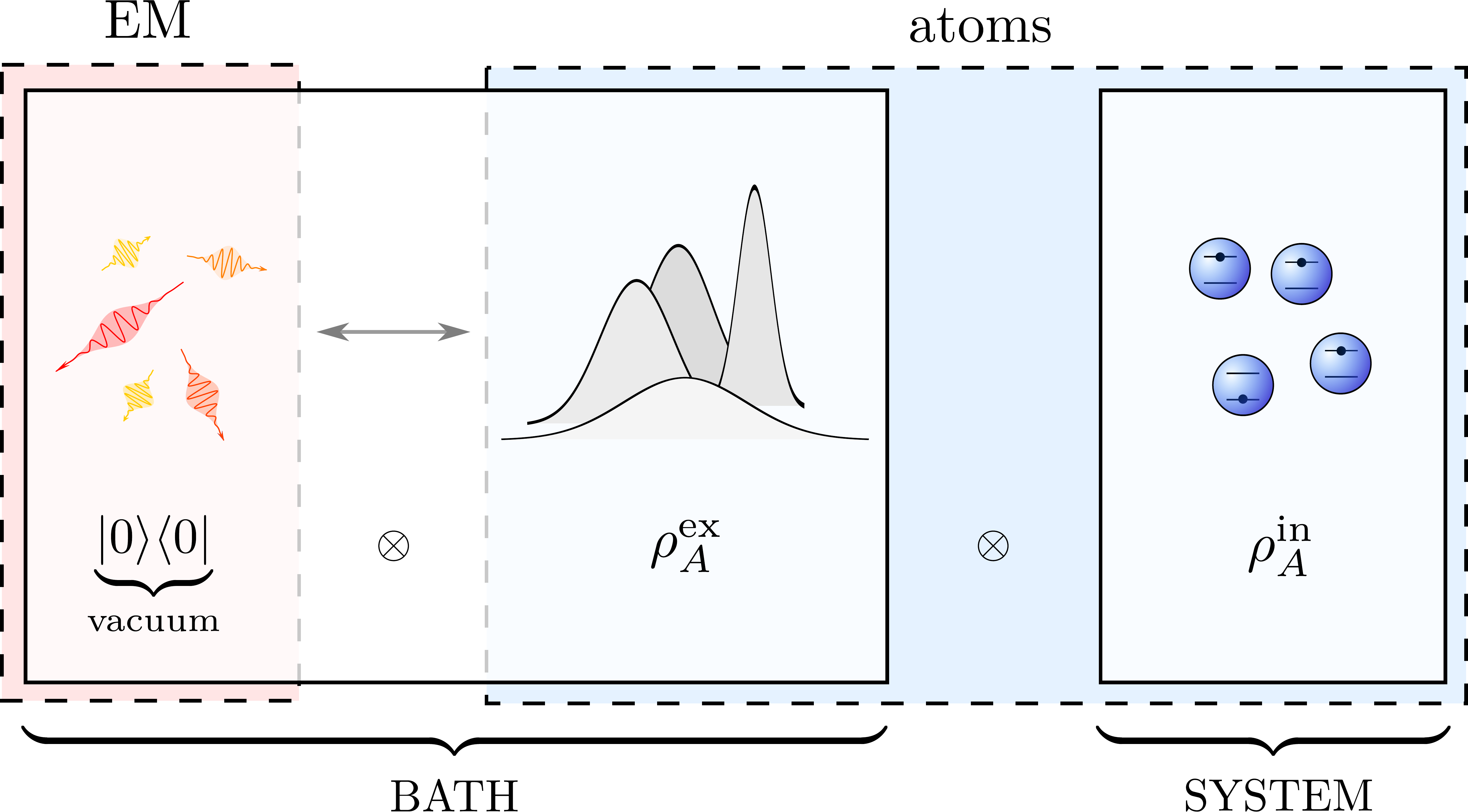}
\end{center}
\caption{(Color online) Decomposition of the global system into bath and system of interest. The system of interest is the internal part of the atoms described by the state $\rho_A^\mathrm{in}$. The bath corresponds to the atomic external degrees of freedom, described by the state $\rho_A^\mathrm{ex}$, and the electromagnetic field degrees of freedom, intially in the vacuum state $\ket{0}\bra{0}$.} \label{system_bath}
\end{figure}

We are interested in the internal dynamics of the atoms only, since our aim is to quantify the effects of the quantization of the atomic motion on cooperative spontaneous emission. In this Section, we derive a Markovian master equation for the internal degrees of freedom from a microscopic approach~\cite{Bre06}. The derivation of a quantum optical master equation is commonly made for atoms at fixed positions. Here, we go beyond this approximation by treating the atomic position quantum mechanically.
The atomic internal degrees of freedom specify our system $S$, and all other degrees of freedom (atomic external and electromagnetic field degrees of freedom) specify the bath $B$ to which $S$ is coupled, as illustrated in Fig.~\ref{system_bath}. 

\subsection{Microscopic derivation and general form of the master equation}

Our starting point is the Liouville-von Neumann evolution
equation
\begin{equation}\label{liou}
i\hbar\frac{d \rho(t)}{dt}=[H_A^{\mathrm{self}}(t) + H_{AF}(t),\rho(t)]
\end{equation}
for the global density matrix $\rho(t)$ in the interaction picture with respect to $H_0$. Time-integration of Eq.~(\ref{liou}) together with a Born series expansion to second order in $H_{AF}$ yields, after tracing over the bath degrees of freedom,
\begin{equation}\label{LvN}
\begin{aligned}
\frac{d
  \rho_A^\mathrm{in}(t)}{dt} &= -\frac{i}{\hbar}\tr_B\left([H_A^{\mathrm{self}}(t),\rho(t)]\right)-\frac{i}{\hbar}\tr_B\left([H_{AF}(t),\rho(0)]\right) \\
  &\hspace{0.4cm} -\frac{1}{\hbar^2}\int_0^t\tr_B\left([H_{AF}(t),[H_{AF}(t'),\rho(t')]]\right)dt'  
  \end{aligned}
\end{equation}
where 
\begin{equation}
\rho_A^\mathrm{in}(t)=\tr_B[\rho(t)]
\end{equation} 
is the reduced density matrix of $S$ (in the interaction picture) describing the atomic internal dynamics.

\subsubsection{Born approximation}

We consider the weak coupling regime and resort to the Born approximation (see e.g.~\cite{Bre06}), which assumes the form
\begin{equation}\label{bornapp}
\rho(t) \approx \rho_A^\mathrm{in}(t) \otimes \rho_B,
\end{equation}
for the global density matrix to describe the time evolution of the system $S$ only. Here $\rho_B = \rho_A^\mathrm{ex} \otimes \rho_F$ is the bath density matrix with $\rho_A^\mathrm{ex}$ the motional density matrix and $\rho_F=\ket{0}\bra{0}$ the electromagnetic field density matrix which we take as the vacuum state~\cite{footnote1b}. The Born approximation excludes correlations between external and internal states. In this approximation, the bath is considered as stationary during the whole relaxation dynamics and the influence of the system on the bath is neglected. Accordingly, we consider in this work that the characteristic evolution time $\tau_M$ of the atomic motion is much larger than the relaxation time $\tau_R$ of the system. This condition is met in a wide range of experimental situations where atoms are optically or magnetically trapped. For example, the typical frequency $\Omega_M$ of a harmonic potential produced with visible light is in the range $1-10^{3}$~Hz, which leads to $\tau_M \sim 1/\Omega_M\gg\tau_R \sim 1/\gamma_0$ where $\gamma_0$ is the single-atom free spontaneous emission rate, of the order of $10^{9}$ Hz for optical transitions (i.e.~there are at least six orders of magnitude separation between $\tau_M$ and $\tau_R$).

In Eq.~(\ref{LvN}), we can furthermore assume without loss of generality that the second term
on the right-hand side vanishes~\cite{footnote2},
which leads to
\begin{equation}\label{LvNB}
\begin{aligned}
\frac{d\rho_A^\mathrm{in}(t)}{dt} &= -\frac{i}{\hbar}\big[\langle H_A^{\mathrm{self}}(t)
\rangle_{\mathrm{ex}},\rho_A^\mathrm{in}(t)\big]
 \\[5pt]
 &\hspace{-0.5cm} -\frac{1}{\hbar^2} \int_0^t  \tr_B([H_{AF}(t),[H_{AF}(t'),\rho_A^\mathrm{in}(t')\otimes\rho_B]])\, dt',
\end{aligned} 
\end{equation}
since $
\tr_B\left([H_A^{\mathrm{self}}(t),\rho(t)]\right) = [\langle H_A^{\mathrm{self}}(t)
\rangle_{\mathrm{ex}},\rho_A^\mathrm{in}(t)]$, where $\langle \,  \cdot \,
\rangle_{\mathrm{ex}}=\mathrm{Tr}(\,\cdot\,\rho_A^{\mathrm{ex}})$
stands for the expectation value over the atomic external degrees of
freedom.

\subsubsection{Markov approximation}
The next step is to perform the Markov approximation to eliminate memory effects and end up with a time-local master equation for $\rho_A^\mathrm{in}(t)$. This can be achieved by making the change of variable $t'\to t-t'$, extending the integration domain to infinity, and replacing $\rho_A^\mathrm{in}(t-t')$ by $\rho_A^\mathrm{in}(t)$ under the integral. This approximation is justified as long as the bath correlation time $\tau_B$ is much smaller than the typical relaxation time $\tau_R$ of the system. It is well established that the Markov approximation is an excellent approximation for describing the process of spontaneous emission of photons from atoms at fixed positions~\cite{Gar04}. We now show that this is also the case when the bath operators $B_j$ [Eq.~(\ref{Bj})] contain in addition the motional degrees of freedom. Inserting Eq.~(\ref{HIgen}) into Eq.~(\ref{LvNB}) yields
\begin{widetext}
\begin{equation}\label{LvNBM}
\frac{d\rho_A^\mathrm{in}(t)}{dt} =-\frac{i}{\hbar}\big[\langle  H_A^{\mathrm{self}}(t)
\rangle_{\mathrm{ex}},\rho_A^\mathrm{in}(t)\big]\, + \sum_{i,j = 1}^{N} \sum_{\omega, \omega' \atop = \pm \omega_0} \Bigg[\Gamma_{ij}(\omega)\,e^{i (\omega'-\omega)t} \left( A^{\mathrm{in}}_j(\omega) \rho_A^\mathrm{in}(t) A_i^{\mathrm{in}\dagger}(\omega') - A_i^{\mathrm{in}\dagger}(\omega') A^{\mathrm{in}}_j(\omega)\rho_A^\mathrm{in}(t) \right) + \mathrm{h.c.} \Bigg],
\end{equation}
\end{widetext}
with the spectral correlation tensor
\begin{equation}\label{eq:corrtensor}
\Gamma_{ij}(\omega)=\frac{1}{\hbar^2}\int_0^\infty e^{i \omega t} \, \cor\,dt,
\end{equation} 
and  the bath correlation function 
\begin{equation}\label{eq:corr}
\cor = \langle B_i^\dagger(t)B_j(0)\rangle_B
\end{equation} 
where $B_j(t)$ is given by Eq.~(\ref{Bj}) and the expectation value is over the bath degrees of freedom. The bath correlation function $\cor$ decays on a time scale $\tau_B$, which defines the bath correlation time. The standard case of atoms at fixed classical positions is obtained formally through the substitution $\hat{\mathbf{r}}_i(t) \to \mathbf{r}_i$ in Eq.~(\ref{Bj}). The correlation function then reduces to $\cor =\langle \mathbf{E}(\br_{i},t)\boldsymbol{\cdot}\mathbf{d}^*\,\mathbf{E}(\br_{j},0)\boldsymbol{\cdot}\mathbf{d}\rangle$ for the electric field components along $\mathbf{d}$. The bath correlation time $\tau_B$ is smaller than an optical period, and thereby much smaller than the spontaneous emission time $\tau_R$ and justifies the Markov
approximation. This is true for both the diagonal ($i=j$) and off-diagonal ($i\ne j$) terms, for all positions $\mathbf{r}_{i}$ and
$\mathbf{r}_{j}$. One might wonder if the motional degrees of freedom induce correlations on a much longer time scale. The relevant bath correlation function $\cor =\langle \mathbf{E}(\hat{\br}_{i}(t),t)\boldsymbol{\cdot}\mathbf{d}^*\,\mathbf{E}(\hat{\br}_{j}(0),0)\boldsymbol{\cdot}\mathbf{d}\rangle$ is still given by the correlation of the field components --- now taken in general at different positions, which are themselves subject to quantum fluctuations and dynamics. However, since the electric field correlations decay on a time scale $\tau_B$ regardless of the positions, we see that the motion of the atoms does not increase the bath correlation time, and the Markov approximation remains therefore justified. 

\subsubsection{Rotating Wave Approximation}
\begin{figure}
\begin{center}
\includegraphics[width=8.5cm]{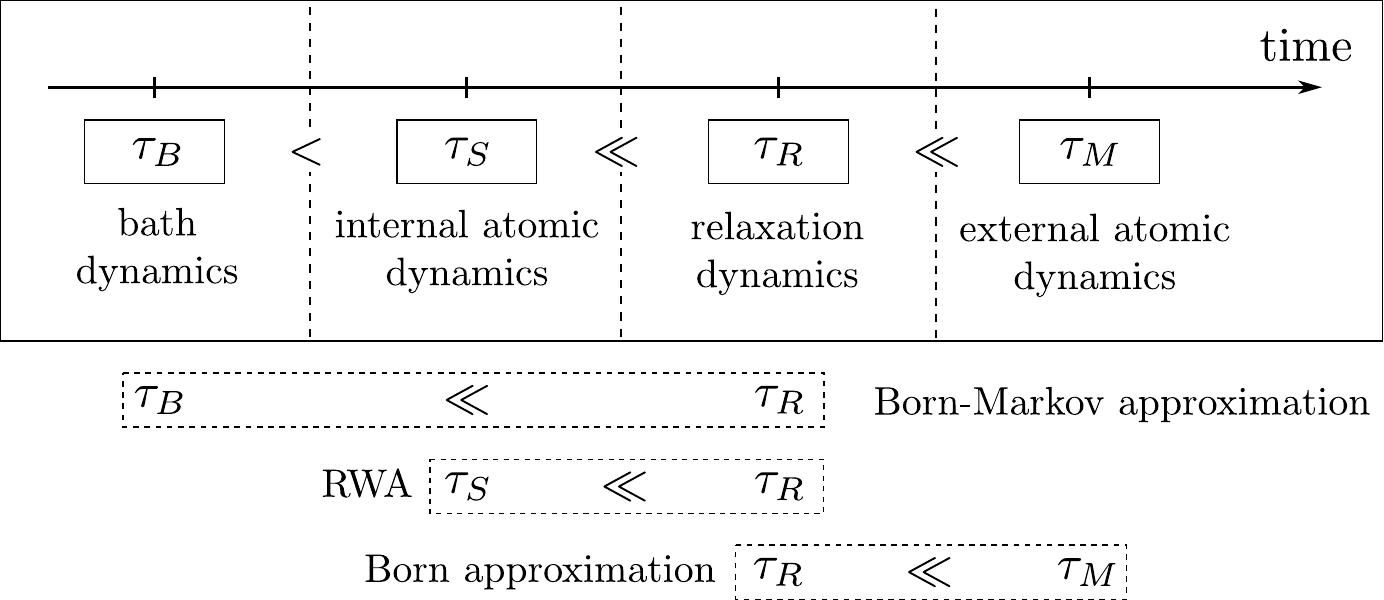}
\end{center}
\caption{Characteristic time scales corresponding to the evolution of the external dynamics ($\tau_M$), the internal dynamics ($\tau_R\sim 1/\gamma_0$ with $\gamma_0$ the free spontaneous emission rate), the isolated system dynamics ($\tau_S\sim 1/\omega_0$ with $\omega_0$ the atomic transition frequency), and the bath ($\tau_B<\tau_S$).} \label{time_scales}
\end{figure}

We now resort to a rotating wave approximation (RWA) by keeping in Eq.~(\ref{LvNBM})
only the energy-conserving terms ($\omega' = \omega$). This ensures that the master equation preserves the positivity of $\rho_A^\mathrm{in}(t)$. The RWA is valid as long as the
relaxation time of the system, $\tau_R\sim 1/\gamma_0$, is much larger than the typical time
scale $\tau_S$ of its intrinsic evolution. Here the intrinsic evolution
corresponds to the internal dynamics of the atoms, hence $\tau_S\sim
1/\omega_0$. We thus have $\tau_S/\tau_R\sim \gamma_0/\omega_0\sim \alpha
(a_0/\lambda_0)^2$ with $\alpha$ the fine-structure constant, $a_0$ the Bohr radius and $\lambda_0$ the wavelength of the emitted radiation. In the optical domain, this 
ratio is much smaller than one and the dipole approximation and RWA are entirely justified. Figure~\ref{time_scales} summarizes all the approximations performed in the derivation of the master equation in terms of the relevant characteristic time scales.

\subsubsection{Correlation functions}\label{seccorfun}

The bath correlation function $\cor$ [Eq.~(\ref{eq:corr})] can be further specified by evaluating the expectation value of the electromagnetic field degrees of freedom. Since the electromagnetic field is initially in vacuum, only the $a_{\keps} a_{\keps}^\dagger$ term survives and the bath correlation function becomes
\begin{equation}\label{eq:corrdec}
\cor = \frac{1}{L^3}\sum_{\keps} \coremk \, \corkm 
\end{equation}
with
\begin{equation}\label{cijclas}
\coremk = \frac{\hbar \omega_k}{2\epsilon_0} \, |\epsk\boldsymbol{\cdot} \mathbf{d}|^2\, e^{- i\omega_k t}
\end{equation}
and the motional correlation function
\begin{equation}\label{eq:com}
\corkm= \big\langle e^{i\mathbf{k}\boldsymbol{\cdot} \hat{\mathbf{r}}_{i}(t)}e^{-i \mathbf{k}\boldsymbol{\cdot} \hat{\mathbf{r}}_{j}(0)}\big\rangle_{\mathrm{ex}}.
\end{equation}
The motional correlation function~(\ref{eq:com}) explicitely
depends on time. However, as explained above, the motion of the atoms
does not increase the bath correlation time. Moreover, since the
typical relaxation time of the internal dynamics, $\tau_R$, is much
smaller than the intrinsic evolution time associated with the atomic
motion, $\tau_M$, the latter is approximately frozen during the
emission of photons, so that $\corkm\approx
\mathcal{C}_{ij}^\mathrm{ex}(\mathbf{k}, 0) \equiv \corkmz$ (see the
discussion after Eq.~\eqref{bornapp}, where we found that the $\tau_M$ and
$\tau_R$ are separated by at least $6$ orders of magnitude in the
typical optical regime). The bath
correlation function then simplifies to 
\begin{equation}
\label{eq:corr2}
 \cor \approx \frac{1}{L^3}\sum_{\keps} \coremk \, \corkmz
\end{equation}
with
\begin{equation}\label{cijex}
\corkmz = \left\langle e^{i\mathbf{k}\boldsymbol{\cdot} \hat{\mathbf{r}}_{ij}}\right\rangle_{\mathrm{ex}} = \mathrm{Tr}_{ij}\left[ e^{i\mathbf{k}\boldsymbol{\cdot} \hat{\mathbf{r}}_{ij}} \rho_{ij}^\mathrm{ex} \right]
\end{equation}
with $\hat{\mathbf{r}}_{ij} = \hat{\mathbf{r}}_{i} - \hat{\mathbf{r}}_j$ and where the trace is now performed over the motional degrees of freedom of the atoms $i$ and $j$ with $\rho_{ij}^\mathrm{ex} $ their external reduced density matrix.

For classical atomic positions, $\hat{\mathbf{r}}_j$ can be replaced by $\mathbf{r}_j$ for all $j$ and the motional correlation function (\ref{cijex}) reduces to $\corkmz=e^{i \mathbf{k}\boldsymbol{\cdot} \mathbf{r}_{ij}}$ with $\mathbf{r}_{ij} = \mathbf{r}_i - \mathbf{r}_j$ the vector connecting atoms $i$ and $j$, so that Eq.~(\ref{cijclas}) yields the Fourier components of the classical correlation function for the electromagnetic field. In contrast, when the atomic motion is quantized, the plane waves $e^{i \mathbf{k}\boldsymbol{\cdot} \mathbf{r}_{ij}}$ in the Fourier series (\ref{eq:corr2}) are replaced by $\langle e^{i\mathbf{k}\boldsymbol{\cdot} \hat{\mathbf{r}}_{ij}}\rangle_{\mathrm{ex}}$ to account for the fluctuations and correlations in the positions of atoms $i$ and $j$.

\subsubsection{Standard form of the master equation}
Under Born-Markov approximation and RWA, Eq.~(\ref{LvNBM}) takes the Lindblad form 
\begin{multline}\label{sf}
\frac{d\rho_A^\mathrm{in}(t)}{dt} =-\frac{i}{\hbar}\big[\langle H_A^{\mathrm{self}} 
\rangle_{\mathrm{ex}},\rho_A^\mathrm{in}(t)\big] \\
- \frac{i}{\hbar}\left[H_\mathrm{\Omega} , \rho_A^\mathrm{in}(t) \right] + \mathcal{D}\left(\rho_A^\mathrm{in}(t)\right)
\end{multline}
with the level-shift Hamiltonian
\begin{equation}\label{hamilcons}
H_\mathrm{\Omega} = \sum_{i, j=1}^{N} \hbar \Omega_{ij} \, \sigma_+^{(i)}\sigma_-^{(j)}
\end{equation}
in terms of level shifts
\begin{equation} 
 \label{Omegaij}
\Omega_{ij} =\mathrm{Im}\left[\Gamma_{ij}(\omega_0)+\Gamma_{ij}(-\omega_0)\right],
\end{equation}
and the dissipator
\begin{equation}
\label{Dissip0}
\mathcal{D}\left(\cdot\right) = \sum_{i,j=1}^N\gamma_{ij}\left(\sigma_-^{(j)}\cdot\sigma_+^{(i)}-\frac{1}{2}\left\{\sigma_+^{(i)}\sigma_-^{(j)},\cdot\right\}\right)
\end{equation}
in terms of decay rates
\begin{equation} 
\label{decaycoeff}
\gamma_{ij} = 2\,\mathrm{Re}\left[\Gamma_{ij}(\omega_0)\right].
\end{equation}
Note that in Eq.~(\ref{sf}), $H_A^\mathrm{self}$ does not depend anymore on time because of the approximation $\hat{\mathbf{r}}_j(t) \approx \hat{\mathbf{r}}_j$ performed above.

Equations~(\ref{hamilcons}) and (\ref{Dissip0}) describe respectively the conservative and dissipative dynamics of the atomic internal state caused by the interaction with the electromagnetic field. The level shifts $\Omega_{ij}$ and the decay rates $\gamma_{ij}$ are obtained from the imaginary and real parts of the spectral correlation tensor $\Gamma_{ij}$ [Eq.~(\ref{eq:corrtensor})]. In the following, we analyse more precisely the structure of these coefficients entering the master equation.

\subsection{Dissipative part}

An explicit expression for the decay rates $\gamma_{ij}$ [Eq.~(\ref{decaycoeff})] can be obtained by performing the time integration in Eq.~(\ref{eq:corrtensor}) together with Eq.~(\ref{eq:corr2}) for the bath correlation function, thereby yielding
\begin{equation}\label{gij2}
\gij = \frac{1}{L^3}\sum_{\keps} \gijemk \, \corkmz
\end{equation}
with
\begin{equation}
\gijemk = \frac{\pi \omega_k}{\hbar \epsilon_0} \,|\epsk\boldsymbol{\cdot}\mathbf{d}|^2\, 
   \delta(\omega_k - \omega_0)
\end{equation}
the Fourier components of the decay rates for classical atomic positions. Equation~(\ref{gij2}) shows that the Fourier components of the decay rates are affected by the quantization of the atomic motion through weighting by the motional correlation function (\ref{cijex}). In the limit of a continuum of modes, the sum over the wave vectors can be replaced by an integral (we use the standard spherical coordinates $(k,\theta,\varphi)$ with $d\Omega = \sin\theta \,d\theta\,d\varphi $),
\begin{equation}
\frac{1}{L^3}\sum_{\mathbf{k}}\to \int \frac{d\mathbf{k}}{(2\pi)^3}\equiv \frac{1}{(2\pi)^3c^3}\int_{0}^{+\infty}\omega^2\,d\omega\int d\Omega,\label{sumtoint}
\end{equation}
and Eq.~(\ref{gij2}) yields, after performing the $\omega$-integration,
\begin{equation}\label{gijgen}
\gij =  \int \sum_{\eps} \gijemkz \,\, \corkzmz  \,\frac{d\Omega}{(2\pi)^2}
\end{equation}
with $\mathbf{k}_0= k_0\,(\cos\varphi\sin\theta, \sin\varphi\sin\theta, \cos\theta)$, $k_0=\omega_0/c$, 
\begin{equation}\label{gijclass}
\gijemkz = \frac{3 \pi \gamma_0}{2} |\boldsymbol{\varepsilon}_{\mathbf{k}_0}\boldsymbol{\cdot}\mathbf{e}_\mathbf{d}|^2
\end{equation}
with $\mathbf{e}_\mathbf{d}=\mathbf{d}/d$, $d=|\mathbf{d}|$ and $\gamma_0$ the single-atom spontaneous emission rate
\begin{equation}\label{saper}
\gamma_0 = \frac{\omega_0^3 d^2}{3\pi\hbar\epsilon_0c^3}.
\end{equation}

For classical atomic positions, $\mathcal{C}_{ij}^{\mathrm{ex}}(\mathbf{k})= e^{i \mathbf{k} \boldsymbol{\cdot} \mathbf{r}_{ij}}$ and Eq.~(\ref{gijgen}) reduces to the classical form of the decay rates for atoms separated by a distance $r_{ij}=|\mathbf{r}_{ij}|$, $\gamma_{ij}=\gamma^{\mathrm{cl}}(\mathbf{r}_{ij})$ with~\cite{Ste64,Aga74}
\begin{equation}
\label{gammaijcl}
\gamma^{\mathrm{cl}}(\mathbf{r}_{ij}) =\frac{3  \gamma_0 }{2}\Bigg[ p_{ij} \,\frac{\sin \xi_{ij}}{\xi_{ij}} + q_{ij}  \left( \frac{\cos \xi_{ij}}{\xi_{ij}^2} - \frac{\sin \xi_{ij}}{\xi_{ij}^3}\right)\Bigg].
\end{equation}
with $\xi_{ij}=k_0 r_{ij}$.
For a $\pi$ transition, the angular factors $p_{ij}$ and $q_{ij}$ are given by
\begin{equation}\label{pqpi}
p_{ij}=\sin^2 \alpha_{ij} ,\;\;  q_{ij}= (1-3 \cos^2 \alpha_{ij}),
\end{equation} 
and for a $\sigma^\pm$ transition by
\begin{equation}\label{pqsigma}
p_{ij}=\tfrac{1}{2}(1+\cos^2 \alpha_{ij}) ,\;\;  q_{ij}=  \tfrac{1}{2}(3 \cos^2 \alpha_{ij}-1)
\end{equation}
with $\alpha_{ij}=\arccos(\mathbf{r}_{ij}\boldsymbol{\cdot}\mathbf{e}_z/r_{ij})$ the angle between the quantization axis and the vector connecting atoms $i$ and $j$. Equation~(\ref{gijgen}) can also be written as
$\gamma_{ij}= \mathcal{F}_{\mathbf{0}}^{-1}\left[ \mathcal{F}_\mathbf{k}\left[ \gamma^{\mathrm{cl}}\right]\corkmz \right]$
which can be seen to be the convolution product $ \left(\gamma^{\mathrm{cl}} \, \star \, f_{ij} \right)(\mathbf{0})$ with $f_{ij}(\mathbf{r} ) = \mathcal{F}^{-1}_{\mathbf{r}}\left[\corkmz \right]$~\cite{footnote3}. Therefore, the decay rates takes the alternative form
\begin{equation}\label{gammaijconv}
\begin{aligned}
\gamma_{ij} &= \int_{\mathbb{R}^3} \gamma^{\mathrm{cl}}(\mathbf{r}) \,\mathcal{F}^{-1}_{\mathbf{r}} \left[\corkmz\right] d\mathbf{r}
\end{aligned}
\end{equation}
in terms of their classical expression~(\ref{gammaijcl}) and the inverse Fourier transform of the motional correlation function~(\ref{cijex}).

Two important features follow from Eq.~(\ref{gijgen}) (or equivalently from Eq.~(\ref{gammaijconv})). First, the diagonal decay rates $\gamma_{ii}$ are seen to coincide with those obtained in the classical case because $\mathcal{C}^\mathrm{ex}_{ii}(\mathbf{k})= 1$ for any motional state and wave vector $\mathbf{k}$. Hence, the dissipative internal dynamics of a single atom is not affected by its motional state when the electromagnetic field is initially in vacuum. Second, Eq.~(\ref{gijgen}) shows that as soon as the quantum nature of the atomic motion becomes appreciable, we have the additional
possibility of influencing the decay rates through engineering the motional
state of the atoms.  The motional correlation function $\corkzmz$ can be seen from
Eq.~(\ref{gijgen}) to play a similar role as mode-dependent modifications of the coupling constants, and can thus be expected to lead to similar effects as Purcell's enhancement or reduction of spontaneous emission~\cite{San11}. 

It readily follows from Eq.~(\ref{gammaijconv}) that $\gamma_{ij}=\gamma_{ji}$ and $|\gamma_{ij}| \leqslant \gamma_0$. Indeed, the classical expression~(\ref{gammaijcl}) satisfies $|\gamma^\mathrm{cl}(\mathbf{r})| \leqslant \gamma^\mathrm{cl}(\mathbf{0}) = \gamma_0 \,\, \forall\, \mathbf{r}$, which implies 
\begin{equation}
\begin{aligned}
|\gamma_{ij}| &\leqslant \gamma_0 \left| \int_{\mathbb{R}^3} \,\mathcal{F}^{-1}_{\mathbf{r}} \left[\corkmz\right] d\mathbf{r} \right|= \gamma_0 \left| \mathcal{C}_{ij}^\mathrm{ex}(\mathbf{0}) \right|=\gamma_0
\end{aligned}
\end{equation}
since $\mathcal{C}_{ij}^\mathrm{ex}(\mathbf{0}) = \mathrm{Tr}(\rho_{ij}^\mathrm{ex}) = 1 $ for any $i,j$ due to normalization.

\subsection{Conservative part}

An explicit expression for the level shifts $\Omega_{ij}$ [Eq.~(\ref{Omegaij})] can be obtained along the same lines as for the decay rates, and reads
\begin{equation}\label{Oijsum}
\Oij = \frac{1}{L^3}\sum_{\keps} \Oijemk \, \corkmz \,
\end{equation}
with
\begin{equation}
\Oijemk =   -\frac{1}{\hbar \epsilon_0}\, 
  \, \mathrm{v.p.}\left(\frac{\omega_k^2}{\omega^2_k-\omega^2_0}\right) |\epsk\boldsymbol{\cdot}\mathbf{d}|^2
\end{equation}
where v.p.\ stands for the Cauchy principal value~\cite{footnote4}. 
In the limit of a continuum of modes, Eq.~(\ref{Oijsum}) becomes
\begin{equation} \label{omegaij}
\Omega_{ij} = \mathrm{v.p.}\int \sum_{\eps} \Oijemk \,
 \corkmz\,\frac{d\mathbf{k}}{(2\pi)^3} 
\end{equation}
with
\begin{equation}
\Oijemk = -  \frac{3 \pi \gamma_0}{k_0^3}\, 
   \frac{k^2}{k^2-k_0^2} \, |\epsk\boldsymbol{\cdot} \mathbf{d}|^2.
\end{equation}
As for the decay rates, the plane waves $e^{i \mathbf{k}\boldsymbol{\cdot} \mathbf{r}_{ij}}$ in the Fourier series for the level shifts $\Omega_{ij}$ are replaced by the motional correlation function (\ref{cijex}) taking into account the quantization of the atomic motion. The diagonal coefficients $\Omega_{ii}$ related to the Lamb shifts are not affected by the quantization of the motion since $\mathcal{C}_{ii}^\mathrm{ex}(\mathbf{k})=1$; they are all equal and can be discarded by means of a renormalization of the atomic transition frequency $\omega_0$. The off-diagonal shifts $\Omega_{ij}$ ($i\ne j$) contain divergent terms, that are already present without quantization of the atomic motion, i.e.~with classical atomic positions. However, these terms are exactly cancelled by other divergent terms appearing in the Hamiltonian $H_A^{\mathrm{self}}$~\cite{Aga74}. This cancellation still holds when the atomic motion is quantized, as we proceed to show. We start by rewriting the Hamiltonian $H_A^{\mathrm{self}}$ [Eq.~(\ref{tHAself})] using the expression of the Dirac delta distribution in integral form in momentum space,
\begin{equation}
\begin{aligned}
H_{A}^{\mathrm{self}} & = \frac{d^2}{2 \epsilon_0} \sum_{i,j=1}^N  \sigma_x^{(i)} \sigma_x^{(j)}\\ 
& \;\;\iiint e^{i (\mathbf{k} -\mathbf{k}')  \boldsymbol{\cdot} \mathbf{r}} e^{-i (\mathbf{k}\boldsymbol{\cdot} \hat{\mathbf{r}}_i - \mathbf{k}' \boldsymbol{\cdot} \hat{\mathbf{r}}_j)}  d\mathbf{r} \,\frac{d\mathbf{k}}{(2\pi)^3}\, \frac{d\mathbf{k}'}{(2\pi)^3} . 
\end{aligned}
\end{equation}
The integration over $\mathbf{r}$ yields a Dirac delta distribution $\delta(\mathbf{k}-\mathbf{k}')$, which eventually leads to the contact interaction Hamiltonian
\begin{equation}\label{Hselfcontact}
H_{A}^{\mathrm{self}}  = \frac{d^2}{2 \epsilon_0} \sum_{i,j=1}^N \sigma_x^{(j)} \sigma_x^{(i)} \delta(\hat{\mathbf{r}}_i - \hat{\mathbf{r}}_j).
\end{equation}
By keeping only the energy conserving terms (RWA) in Eq.~(\ref{Hselfcontact}), the expectation value $\langle H_A^{\mathrm{self}} 
\rangle_{\mathrm{ex}}$ appearing in Eq.~(\ref{sf}) becomes
\begin{equation}
\begin{aligned}\label{selfh2}
\langle H_A^{\mathrm{self}} 
\rangle_{\mathrm{ex}} &= \sum_{i \neq j}^N
\hbar \Omega_{ij}^{\mathrm{self}} \sigma_+^{(j)} \sigma_-^{(i)} + \sum_{i=1}^N 
\hbar \Omega_{ii}^{\mathrm{self}} \,\mathbb{1}^{(i)}
\end{aligned}
\end{equation}
with $\mathbb{1}^{(i)}$ the internal identity operator for atom $i$ and 
\begin{align}\label{omegaijself}
&\Omega_{ij}^\mathrm{self} = \frac{3\pi \gamma_0}{k_0^3} \int \corkmz\,\frac{d\mathbf{k}}{(2\pi)^3}  , \\
&\Omega_{ii}^\mathrm{self} = \frac{3\pi \gamma_0}{2 k_0^3} \int \frac{d\mathbf{k}}{(2\pi)^3}.
\end{align}
Since the divergent level-shift $\Omega_{ii}^\mathrm{self}$ in Eq.~(\ref{selfh2}) is proportional to the identity, it can be absorbed by means of a redefinition of the zero energy, so that $\langle H_A^{\mathrm{self}} 
\rangle_{\mathrm{ex}}$ reduces to
\begin{equation}
\langle H_A^{\mathrm{self}} 
\rangle_{\mathrm{ex}} = \sum_{i \neq j}^N
\hbar \Omega_{ij}^{\mathrm{self}} \sigma_+^{(j)} \sigma_-^{(i)}.
\end{equation}

We now split the level shifts $\Omega_{ij}$ [Eq.~(\ref{omegaij})] into~\cite{Aga74}
\begin{equation}
\Omega_{ij}= \Delta_{ij} - \Omega_{ij}^\mathrm{self}
\end{equation}
where $\Omega^{\mathrm{self}}_{ij}$ is given by Eq.~(\ref{omegaijself}) and $\Delta_{ij}$ is the dipole-dipole shift given by
\begin{equation}\label{deltaijgen}
\Delta_{ij} = \mathrm{v.p.}\int\sum_{\eps} \Dijemk \, \corkmz\,\frac{d\mathbf{k}}{(2\pi)^3} 
\end{equation}
with
\begin{equation}\label{Dijclkc}
\Dijemk = \frac{3\pi \gamma_0}{k_0^3}  \left[ 1 -  \frac{k^2}{k^2-k^2_0} |\epsk\boldsymbol{\cdot}
\mathbf{e}_\mathbf{d}|^2 \right].
\end{equation}

The Hamiltonian (\ref{hamilcons}) entering the master equation can then be decomposed as
\begin{equation}
H_\mathrm{\Omega} =  \sum_{i \neq  j}^{N} \hbar\Omega_{ij} \, \sigma_+^{(i)}\sigma_-^{(j)} \equiv H_\mathrm{\Delta} - \langle H_A^{\mathrm{self}} 
\rangle_{\mathrm{ex}} \label{hdeltaij}
\end{equation}
with the dipole-dipole Hamiltonian
\begin{equation}\label{dipdipHa}
H_\mathrm{\Delta} = \sum_{i \neq j }^{N} \hbar \Delta_{ij} \, \sigma_+^{(i)}\sigma_-^{(j)},
\end{equation}
so that Eq.~(\ref{sf}) eventually reads
\begin{equation}\label{sfsch}
\begin{aligned}
\frac{d\rho_A^\mathrm{in}(t)}{dt} &= - \frac{i}{\hbar}\left[ H_\mathrm{\Delta}, \rho_A^\mathrm{in}(t) \right] + \mathcal{D}\left(\rho_A^\mathrm{in}(t)\right).
\end{aligned}
\end{equation}
Hence, $H_A^\mathrm{self}$ does not contribute to the dynamics, and $H_\mathrm{\Delta}$ is the proper form of the Hamiltonian to describe the conservative dynamics of the atomic system. It accounts for second order photon exchanges between pairs of atoms in different internal energy eigenstates~\cite{footnote5}.

For classical atomic positions, $\mathcal{C}_{ij}^{\mathrm{ex}}(\mathbf{k})= e^{i \mathbf{k} \boldsymbol{\cdot} \mathbf{r}_{ij}}$ and Eq.~(\ref{deltaijgen}) reduces to the retarded interaction energy (divided by $\hbar$) between two parallel dipoles located at fixed positions $\mathbf{r}_{i}$ and $\mathbf{r}_{j}$, i.e.\ $\Delta_{ij}=\Delta^{\mathrm{cl}}(\mathbf{r}_{ij})$ with~\cite{Ste64,Aga74}
\begin{equation}
\label{deltaijcl}
\Delta^{\mathrm{cl}}(\mathbf{r}_{ij}) =\frac{3  \gamma_0 }{4}\Bigg[ - p_{ij} \,\frac{\cos \xi_{ij}}{\xi_{ij}} + q_{ij}  \left( \frac{\sin \xi_{ij}}{\xi_{ij}^2} + \frac{\cos \xi_{ij}}{\xi_{ij}^3}\right) \Bigg],
\end{equation} 
$\xi_{ij}=k_0r_{ij}$ and where $p_{ij}$ and $q_{ij}$ are given by Eq.~(\ref{pqpi}) for a $\pi$ transition, and by Eq.~(\ref{pqsigma}) for a $\sigma^\pm$ transition. The sum over the polarizations of the Fourier components~(\ref{Dijclkc}) is thus equal to the Fourier transform of the retarded dipole-dipole interaction energy (divided by $\hbar$), $\Delta^{\mathrm{cl}}(\mathbf{r})$. Equation~(\ref{deltaijgen}) is the generalization of the dipole-dipole shifts (\ref{deltaijcl}) to account for quantum fluctuations and correlations in the atomic motion. Similarly to the decay rates, the dipole-dipole shifts can be written as 
\begin{equation}\label{deltaijconv2}
\begin{aligned}
\Delta_{ij} &= \int_{\mathbb{R}^3} \Delta^{\mathrm{cl}}(\mathbf{r}) \,\mathcal{F}^{-1}_{\mathbf{r}} \left[\corkmz\right] d\mathbf{r}.
\end{aligned}
\end{equation}
As an example, let us consider again the case of two atoms at classical positions $\mathbf{r}_i$ and $\mathbf{r}_j$. We then have $\mathcal{C}_{ij}^{\mathrm{ex}}(\mathbf{k})= e^{i \mathbf{k}\boldsymbol{\cdot} \mathbf{r}_{ij}}$ and Eq.~(\ref{deltaijconv2}) reduces to $\Delta_{ij}=\Delta^{\mathrm{cl}}(\mathbf{r}_{ij})$, as expected. However, in most cases, Eq.~(\ref{deltaijconv2}) yields an infinite result because the
$1/r^3$ divergence of $\Delta^{\mathrm{cl}}(\mathbf{r})$ at $r=0$ is not integrable in $\mathbb{R}^3$ and because $\mathcal{F}^{-1}_{\mathbf{r}} [\corkmz]$ does in general not vanish at the origin. In order to treat dipole-dipole interactions, one must introduce a minimal distance, i.e.\ a cutoff, in the integral (\ref{deltaijconv2}). A natural cutoff would be of the order of the size of an atom, so as to remain compatible with the dipole approximation made in the derivation of the master equation. The effect of the cutoff will be discussed in detail in the following sections.

\section{General expressions of decay rates and dipole-dipole shifts}
\label{secdecay}

The master equation (\ref{sfsch}) is completely determined in terms of the motional correlation function~(\ref{cijex}) through the expressions of the decay rates $\gamma_{ij}$, given by Eq.~(\ref{gammaijconv}) and appearing in the dissipator (\ref{Dissip0}), and the dipole-dipole shifts $\Delta_{ij}$, given by Eq.~(\ref{deltaijconv2}) and appearing in the dipole-dipole Hamiltonian (\ref{dipdipHa}). All the effects related to recoil, quantum fluctuations of motion and indistinguishability are included in the motional correlation function $\corkmz$. In this section, we provide general expressions for $\corkmz$, $\gamma_{ij}$ and $\Delta_{ij}$ both for distinguishable and indistinguishable atoms for arbitrary motional states.

\subsection{Distinguishable atoms}

When $N$ distinguishable atoms are in the motional separable state $\ket{\phi_{1_{\ell}} \dotsc \phi_{N_{\ell}}}$ with a probability $p_\ell \geqslant 0$ ($\sum_\ell p_\ell = 1$), the global motional state is the statistical mixture
\begin{equation}\label{rhoAexdis}
\rho_A^{\mathrm{ex,sep}} = \sum_{\ell = 1}^{L} p_\ell \, \ket{\phi_{1_{\ell}} \dotsc \phi_{N_{\ell}}}\bra{\phi_{1_{\ell}} \dotsc \phi_{N_{\ell}}}.
\end{equation}
The single-atom motional states $\ket{\phi_{j_{\ell}}}$ ($j=1,\ldots,N$; $\ell=1,\ldots,L$) are normalized but are not necessarily orthogonal. The two-atom reduced density matrix $\rho_{ij}^\mathrm{ex}$ is obtained by tracing over the motional degrees of freedom of all atoms but $i$ and $j$, and reads
\begin{equation}\label{rhoijdis}
\rho_{ij}^{\mathrm{ex,sep}} = \sum_{\ell = 1}^{L}p_\ell  \,  \ket{\phi_{i_{\ell}} \phi_{j_{\ell}}} \bra{\phi_{i_{\ell}}\phi_{j_{\ell}}}.
\end{equation}
The motional correlation function (\ref{cijex}) is thus given, for distinguishable atoms (in the mixture (\ref{rhoAexdis})), by
\begin{equation}\label{cijexdis}
\mathcal{C}_{ij}^\mathrm{\,ex,sep}(\mathbf{k}) = \sum_{\ell = 1}^{L}  p_\ell \, I_{i_{\ell} i_{\ell}}(\mathbf{k}) I_{j_{\ell} j_{\ell}}(-\mathbf{k}) 
\end{equation}
with the overlap integral
\begin{equation}\label{overlap}
\begin{aligned}
I_{\alpha \beta} (\mathbf{k})
&= \int_{\mathbb{R}^3} e^{i \mathbf{k} \boldsymbol{\cdot} \mathbf{r}}  \, \phi_{\alpha}(\mathbf{r}) \, \phi^*_{\beta}(\mathbf{r})\,d\mathbf{r}  \\
&= \int_{\mathbb{R}^3} \mathcal{F}_{\mathbf{k}' - \mathbf{k}}[{\phi_{\alpha}}] \, \mathcal{F}_{\mathbf{k}'}[\phi^{*}_{\beta}]\, d\mathbf{k}'
\end{aligned}
\end{equation}
defined for any pair of indices $\alpha\beta$. The overlap integral (\ref{overlap}) is equal to the overlap in momentum space between the state $\phi_{\beta}$ and the state $\phi_{\alpha}$ shifted by the momentum $\hbar \mathbf{k}$ of a photon of wave vector $\mathbf{k}$. The inverse Fourier transform of (\ref{cijexdis}) can be written 
\begin{equation}\label{FinvCijex}
\mathcal{F}^{-1}_{\mathbf{r}} \left[\corkmz\right] = \sum_{\ell = 1}^L p_\ell \int_{\mathbb{R}^3} |\phi_{i_\ell}(\mathbf{r}')|^2 \,|\phi_{j_\ell}(\mathbf{r}+\mathbf{r}')|^2 \, d\mathbf{r}'.
\end{equation}
On inserting Eq.~(\ref{FinvCijex}) into Eqs.~(\ref{gammaijconv}) and (\ref{deltaijconv2}), we obtain explicit expressions for the decay rates and the dipole-dipole shifts in terms of single-atom motional states
\begin{equation}\label{disEXCHANGE1}
\gamma_{ij}^{\mathrm{sep}} = \sum_{\ell = 1}^L p_\ell \iint_{\mathbb{R}^3\times \mathbb{R}^3} \gamma^{\mathrm{cl}}(\mathbf{r}-\mathbf{r}') \,|\phi_i(\mathbf{r})|^2 \,|\phi_j(\mathbf{r}')|^2 \,d\mathbf{r}\, d\mathbf{r}',
\end{equation}
\begin{equation}\label{disEXCHANGE2}
\Delta_{ij}^{\mathrm{sep}} = \sum_{\ell = 1}^L p_\ell \iint_{\mathbb{R}^3\times \mathbb{R}^3} \Delta^{\mathrm{cl}}(\mathbf{r}-\mathbf{r}') \,|\phi_i(\mathbf{r})|^2 \,|\phi_j(\mathbf{r}')|^2 \,d\mathbf{r}\, d\mathbf{r}'.
\end{equation}

\subsection{Indistinguishable atoms}

For indistinguishable atoms in a statistical mixture $\rho_A$, each wave function of the mixture has to be either symmetric or antisymmetric under exchange of particles, depending on the quantum statistics of the atoms (bosonic or fermionic). Due to the Born approximation, the mixture contains a single term and the initial state has to be of the form $\rho_A(0)=\rho_A^\mathrm{in}\otimes \rho_A^\mathrm{ex}$. For clarity, we shall consider \emph{pure} product initial states, and restrict ourselves to states that are both individually either symmetric ($+$) or antisymmetric ($-$). The symmetrization (antisymmetrization) of the separable motional state $\ket{\phi_{1} \dotsc \phi_{N}}$ leads to the $N$-atom symmetric (antisymmetric) state
\begin{equation}
\begin{aligned}
\ket{\Phi_A^\mathrm{ex,\pm}}=\sqrt{\frac{n_{\phi_1}!\cdots n_{\phi_N}!}{N!}}\,\sum_{\pi}  s_{\pm}^{\pi} \,
\ket{\phi_{\pi(1)}
\cdots \phi_{\pi(N)}}
\end{aligned}
\end{equation}
where $n_{\phi_j}$ is the number of atoms occupying the single-atom motional state $\ket{\phi_j}$, the sum runs over all permutations $\pi$ of the indices $\{1, \dotsc, N\}$, and the symbol $ s_{\pm}^{\pi} $ is defined as
\begin{equation}
 s_{\pm}^{\pi} = \begin{cases} 1 & \mbox{if $+$}, \\ 
 \mathrm{sign}(\pi) & \mbox{if $-$}, \end{cases}
\end{equation}
where $\mathrm{sign}(\pi)$ is the signature of the permutation $\pi$. 
The two-atom reduced density matrix, obtained by taking the partial trace of $\rho_A^{\mathrm{ex, \pm}}=\ket{\Phi_A^\mathrm{ex,\pm}}\bra{\Phi_A^\mathrm{ex,\pm}}$ over all atoms but $i$ and $j$, has the form
\begin{equation}\label{rhoijpm}
\begin{aligned}
\rho_{ij}^{\mathrm{ex, \pm}} &= \sum_{\pi, \pi'}\lambda_{ij}^{\pi\pi', \pm} \,\ket{\phi_{\pi(i)}\phi_{\pi(j)}} \bra{\phi_{\pi'(i)}\phi_{\pi'(j)}}
\end{aligned}
\end{equation}
with
\begin{equation}\label{pij}
\lambda_{ij}^{\pi\pi', \pm} = \frac{\displaystyle  s_{\pm}^{\pi} \, s_{\pm}^{\pi'}  \, \prod_{n = 1 \atop n\neq i,j}^N \langle\phi_{\pi'(n)}|\phi_{\pi(n)}\rangle}{\displaystyle \sum_{\tilde{\pi},\tilde{\pi}'}  s_{\pm}^{\tilde{\pi}} \, s_{\pm}^{\tilde{\pi}'} \prod_{n = 1}^N\langle\phi_{\tilde{\pi}'(n)}|\phi_{\tilde{\pi}(n)}\rangle}.
\end{equation}
Inserting Eq.~(\ref{rhoijpm}) into (\ref{cijex}) eventually leads to the motional correlation function
\begin{equation}\label{cijexpm}
\begin{aligned}
\mathcal{C}_{ij}^\mathrm{\,ex, \pm}(\mathbf{k}) = \sum_{\pi, \pi'} \lambda_{ij}^{\pi\pi',\pm} \, I_{\pi(i)\pi'(i)}(\mathbf{k}) \, I_{\pi(j)\pi'(j)}(- \mathbf{k}).
\end{aligned}
\end{equation}
An important result is that $\mathcal{C}_{ij}^\mathrm{\,ex, \pm}$, and thus $\gamma_{ij}$ and $\Delta_{ij}$ [see Eqs.~(\ref{gijgen}) and (\ref{deltaijconv2})], do not depend on $i$ and $j$ for indistinguishable atoms, regardless of the average distance between atoms. This fact has far reaching consequences on how the atomic system radiates, especially in the regime in which cooperative processes are enhanced, when atoms are located within a volume smaller than $\lambda_0^3$. For distinguishable atoms, cooperative emission (superradiance or subradiance) is strongly altered by the dephasing of the atomic dipoles as a consequence of dipole-dipole interactions, whereas for indistinguishable atoms no such dephasing occurs.

The reduced density matrix~(\ref{rhoijpm}) leads to decay rates and dipole-dipole shifts in terms of the following \emph{exchange integrals}
\begin{multline}
\gamma_{ij} = \sum_{\pi, \pi'} \lambda_{ij}^{\pi\pi',\pm} \iint_{\mathbb{R}^3\times \mathbb{R}^3} \gamma^{\mathrm{cl}}(\mathbf{r}-\mathbf{r}') \,\phi_{\pi(i)}(\mathbf{r}) \phi_{\pi'(i)}^*(\mathbf{r}) \\\label{indisEXCHANGE1}
\times \phi_{\pi(j)}(\mathbf{r}')\phi_{\pi'(j)}^*(\mathbf{r}')
 \,d\mathbf{r}\, d\mathbf{r}',
\end{multline}
\begin{multline}
\Delta_{ij} = \sum_{\pi, \pi'} \lambda_{ij}^{\pi\pi',\pm} \iint_{\mathbb{R}^3\times \mathbb{R}^3} \Delta^{\mathrm{cl}}(\mathbf{r}-\mathbf{r}') \,\phi_{\pi(i)}(\mathbf{r}) \phi_{\pi'(i)}^*(\mathbf{r}) \\ \label{indisEXCHANGE2}
\times \phi_{\pi(j)}(\mathbf{r}')\phi_{\pi'(j)}^*(\mathbf{r}')
 \,d\mathbf{r}\, d\mathbf{r}'.
\end{multline}
A particularly relevant situation in the context of cold-atom physics is when all atoms occupy the same motional state $\ket{\phi_0}$ and thus form a Bose-Einstein condensate, i.e.\ when the global motional state $\rho_A^\mathrm{ex}=(\ket{\phi_0}\bra{\phi_0})^{\otimes N}$ is symmetric and separable. The corresponding correlation function is given by Eq.~(\ref{cijexdis}) for $L = 1$ and can be simplified into
\begin{equation}
\mathcal{C}_{ij}^{\mathrm{ex},+}(\mathbf{k}) = I_{00}(\mathbf{k}) \, I_{00}(-\mathbf{k})=\big|\mathcal{F}_{\mathbf{k}}\left[ \,|\phi_0(\mathbf{r})|^2\, \right]\!\big|^2.
\end{equation}
The decay rates (\ref{indisEXCHANGE1}) and dipole-dipole shifts (\ref{indisEXCHANGE2}) read in this case
\begin{equation}\label{beEXCHANGE1}
\gamma_{ij} = \iint_{\mathbb{R}^3\times \mathbb{R}^3} \gamma^{\mathrm{cl}}(\mathbf{r}-\mathbf{r}') \:|\phi_0(\mathbf{r})|^2 \: |\phi_0(\mathbf{r}')|^2 \,d\mathbf{r}\, d\mathbf{r}',
\end{equation}
\begin{equation}\label{beEXCHANGE2}
\Delta_{ij} = \iint_{\mathbb{R}^3\times \mathbb{R}^3} \Delta^{\mathrm{cl}}(\mathbf{r}-\mathbf{r}') \:|\phi_0(\mathbf{r})|^2 \: |\phi_0(\mathbf{r}')|^2 \,d\mathbf{r}\, d\mathbf{r}'.
\end{equation}

\section{Decay rates and dipole-dipole shifts for particular motional states}

\begin{figure}
\begin{center}
\includegraphics[width=6.5cm]{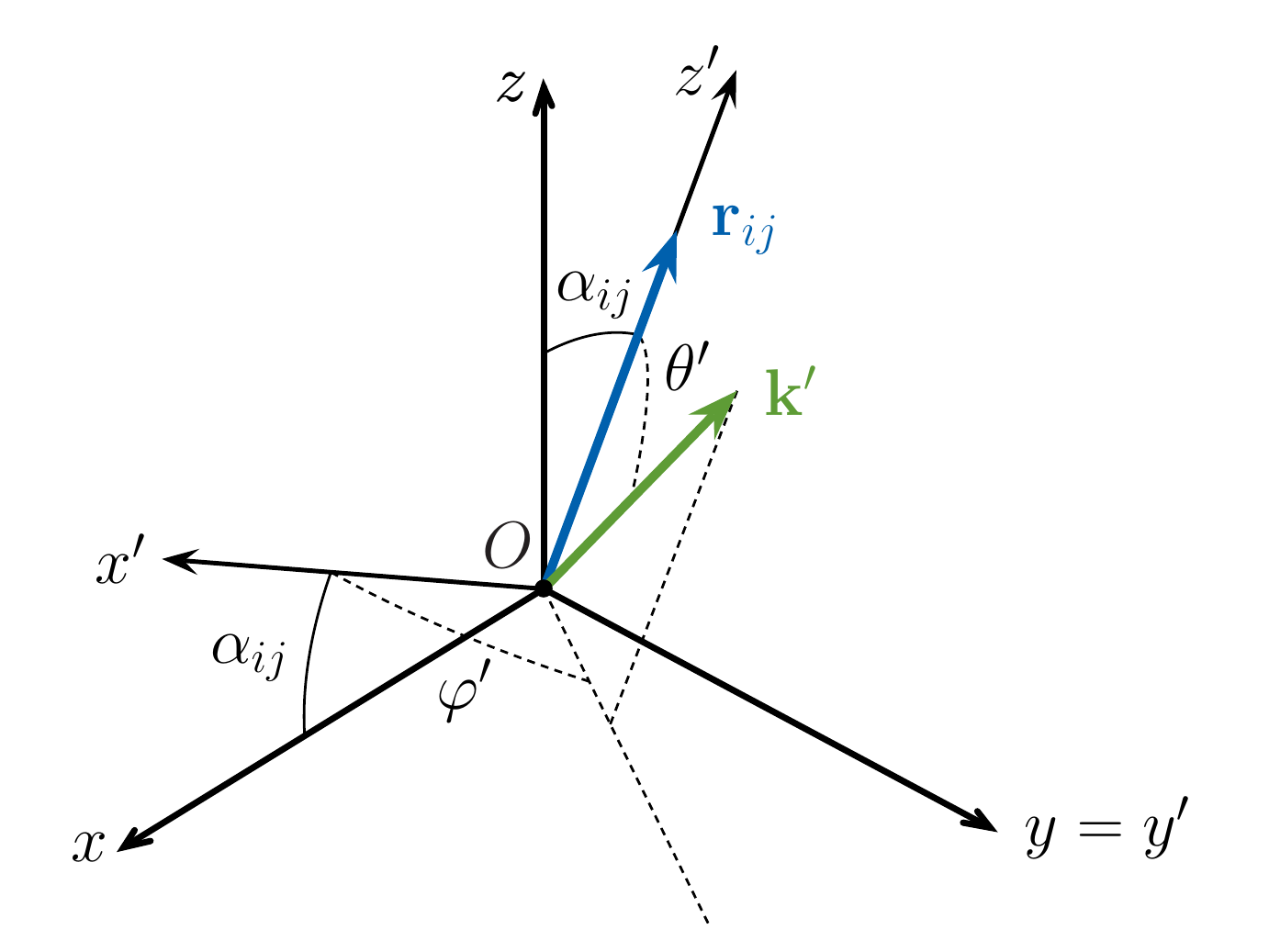}
\end{center}
\caption{(Color online) Coordinate system $Oxyz$ where the $z$-direction corresponds to the quantization axis. The dipole moment for a $\pi$ transition is $\mathbf{d}_\pi=d_0\,\mathbf{e}_z$ with $d_0\in\mathbb{R}$, and for a $\sigma^{\pm}$ transition is $\mathbf{d}_{\sigma^{\pm}}=d_\mp\,\eps_\mp$ with $d_\mp\in\mathbb{C}$. The vector $\mathbf{r}_{ij}$ connecting the atoms $i$ and $j$ lies in the plane $y = 0$ and forms an angle $\alpha_{ij}$ with the $z$-axis. The primed coordinate system $Ox'y'z'$, equiped with spherical coordinates $(k',\theta',\varphi')$, is chosen so that $\mathbf{r}'_{ij}\equiv \mathbf{r}_{ij}$ lies along the $z'$ axis in order to facilitate the calculation of the correlation functions.} \label{coordinate_system}
\end{figure}

In this section, we determine \emph{explicit} expressions for the decay rates $\gamma_{ij}$ and the dipole-dipole shifts $\Delta_{ij}$ for different motional states of particular interest. We also discuss the effects of quantum statistics by considering both cases of distinguishable and indistinguishable atoms. For calculation purposes, it is convenient to work in the coordinate system $Ox'y'z'$ as depicted in Fig.~\ref{coordinate_system} with the $z'$-axis along the vector $\mathbf{r}'_{ij}\equiv \mathbf{r}_{ij}$ connecting the atoms $i$ and $j$, so that $\mathbf{k}'_0\boldsymbol{\cdot}\mathbf{r}'_{ij} = k_0 r_{ij} \cos \theta'$. This coordinate system results from a clockwise rotation of $Oxyz$ by an angle $\alpha_{ij}$ around the $y$ axis.

\subsection{Gaussian states} 

Gaussian wave packets are of particular importance because they describe a broad class of states, such as the ground state of atoms trapped in harmonic potential, realized e.g.\ in a non-interacting Bose-Einstein condensate at zero temperature, but also non-classical states such as squeezed vibrational states of ions in harmonic trap \cite{Hep95}. We consider $N$ single-atom Gaussian wave packets
\begin{equation}\label{Gaussianstate}
\phi_j(\mathbf{r}') \equiv \langle \mathbf{r}' \ket{\phi_j} = \prod_{u = x',y',z'} \sqrt{\frac{1}{\sqrt{2\pi} \sigma_{u}}}\,e^{-(u-u_j)^2/4\sigma_{u}^2}
\end{equation}
for $j = 1, \dotsc, N$. The wave packets are centered around arbitrary positions $\mathbf{r}'_j = (x'_j, y'_j, z'_j)$ with widths $\sigma_{x'}$, $\sigma_{y'}$ and $\sigma_{z'}$ corresponding to the standard deviations along the three spatial directions, taken equal for all atoms. These states can be seen as the ground states of 3D-harmonically trapped atoms, with $\mathbf{r}'_j$ the position of the center of the trap, $\Omega_u=\hbar /2M\sigma_{u}^2$ its frequency along the $u$-direction ($u = x',y',z'$) and $M$ the atomic mass. Plugging Eq.~(\ref{Gaussianstate}) into (\ref{overlap}), we get for the overlap integral between any two Gaussian states $\ket{\phi_i}$ and $\ket{\phi_j}$
\begin{equation}
\begin{aligned}\label{overlapGS}
I_{ij}(\mathbf{k}) &=  \prod_{u = x',y',z'} e^{-k_u\left[k_u \sigma_{u}^2 - i (u_i + u_j)\right]/2}\,e^{-(u_i-u_j)^2/8\sigma_{u}^2}.
\end{aligned}
\end{equation}

For simplicity, we now consider $\sigma_{x'} \to 0$ and $\sigma_{y'} \to 0$, so that the atomic motion is only quantized along the $z'$-direction, hence along $\mathbf{r}'_{ij}=(0,0,z'_{ij})$. This is the most interesting case of quantization along only one direction as it allows for the spatial overlap of atomic wave packets. This choice of coordinate system can always be made for $N=2$ atoms, and the results that we obtain can be transposed to more than two atoms as long as the atoms are aligned along the $z'$-direction. From now on, we denote by $\ell_0\equiv \sigma_{z'}$ the standard deviation of the Gaussian wave packet along this direction.

\subsubsection{Distinguishable atoms}

\begin{figure}
\begin{center}
\includegraphics[width=0.4\textwidth]{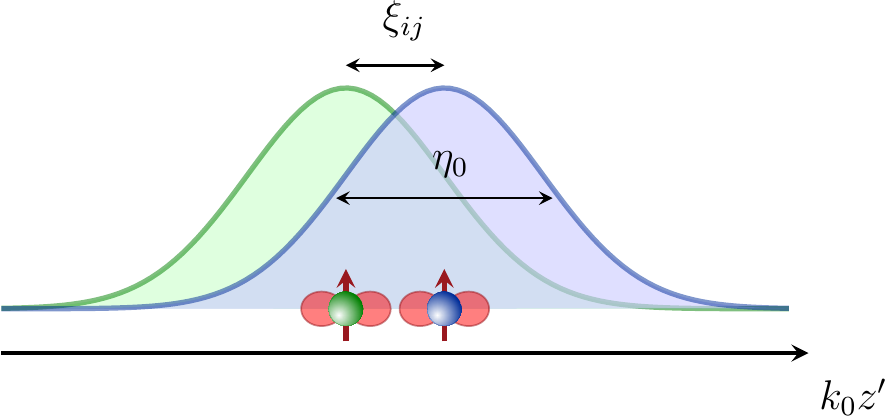}
\end{center}
\caption{(Color online) Schematic view of two atoms separated by a distance $r_{ij}$ and whose external states are described by Gaussian wave packets of width $\ell_0$. The dipole moments and dipole radiation patterns are illustrated in red and correspond to a $\pi$ transition with $\alpha_{ij}=\pi/2$. } \label{fig:gaussians}
\end{figure}

For two distinguishable atoms $i$ and $j$ in the states $\ket{\phi_{i}}$ and $\ket{\phi_{j}}$ respectively, Eq.~(\ref{overlapGS}) yields for the correlation function (\ref{cijexdis})
\begin{align}
\mathcal{C}_{ij}^{\mathrm{ex, sep}}(\mathbf{k}') &= I_{ii}(\mathbf{k}')\,I_{jj}(-\mathbf{k}') \nonumber\\
&= e^{-(k' \ell_0 \cos\theta')^2} e^{i k' r_{ij} \cos\theta'}. \label{corkmGSdis}
\end{align}
In the limit of tight confinement, $\ell_0 \to 0$ , atoms are well localized and the correlation function reduces to its classical expression $e^{i \mathbf{k}' \boldsymbol{\cdot} \mathbf{r}'_{ij}}$. For any other value of $\ell_0$, the decay rates (\ref{gijgen}) resulting from the correlation function (\ref{corkmGSdis}) are obtained from the angular integral
\begin{equation}
\begin{aligned}\label{gijGSintegral}
\gamma_{ij}^\mathrm{sep} &= \frac{3 \gamma_0}{8\pi} \int \sum_{\eps'} |\boldsymbol{\varepsilon}'_{\mathbf{k}'_0}\boldsymbol{\cdot}\mathbf{e}'_{\mathbf{d}'}|^2 e^{-(k_0 \ell_0 \cos\theta')^2}  e^{i k_0 r_{ij} \cos\theta'} d\Omega'
\end{aligned}
\end{equation}
with $d\Omega' = \sin\theta'd\theta'd\varphi'$ and where the sum over the polarizations yields a factor $\sum_{\eps'} |\boldsymbol{\varepsilon}'_{\mathbf{k}'_0}\boldsymbol{\cdot}\mathbf{e}'_{\mathbf{d}'}|^2 = 1 - \mu_{ij}$
with
\begin{equation}\label{polapi}
\mu_{ij} = \left(\cos\varphi' \sin \alpha_{ij} \sin  \theta' + \cos \alpha_{ij} \cos \theta'\right)^2
\end{equation}
for a $\pi$ transition
and
\begin{equation}\label{polapm}
\begin{aligned}
\mu_{ij} = \frac{\left(\cos \theta' \sin \alpha_{ij} - \cos \alpha_{ij} \cos\varphi' \sin\theta' \right)^2 - \sin^2\theta' \sin^2\varphi'}{2}
\end{aligned}
\end{equation}
for a $\sigma^{\pm}$ transition. The integral can be evaluated analytically and provides us with the closed formula
\begin{widetext}
\begin{equation}
\label{gijgsdis}
\gamma^{\mathrm{sep}}_{ij}(\mathbf{r}_{ij},\ell_0) = \frac{3\gamma_0}{16 \eta_0^5} \bigg(\frac{\sqrt{\pi}}{6} e^{-\frac{\xi_{ij}^2}{4\eta_0^2}}\big[ 16 \eta_0^4 - q_{ij}\left(4 \eta_0^4 + 3 \xi_{ij}^2 - 6 \eta_0^2 \right) \big]\mathrm{Re}\left\{ \mathrm{erf}\left( \eta_0 + \frac{i \xi_{ij}}{2 \eta_0} \right) \right\} - q_{ij} \eta_0 e^{-\eta_0^2} \left[2 \eta_0^2 \cos \xi_{ij} - \xi_{ij}\sin \xi_{ij} \right] \bigg)
\end{equation}
\end{widetext}
where $\mathrm{erf}(z)=\frac{2}{\sqrt{\pi }}\int _0^ze^{-t^2}d t $ is the error function, $q_{ij}$ is the angular factor given by Eq.~(\ref{pqpi}) for a $\pi$ transition and by Eq.~(\ref{pqsigma}) for a $\sigma^{\pm}$ transition, and
\begin{equation}
\begin{aligned}\label{LDD}
&\xi_{ij} = k_0 r_{ij} = 2\pi \frac{r_{ij}}{\lambda_0}, \hspace{1cm} \eta_0 = k_0 \ell_0 = 2\pi \frac{\ell_0}{\lambda_0} .
\end{aligned}
\end{equation}
The parameter $\xi_{ij}$ quantifies the significance of atomic cooperative processes. The Lamb-Dicke parameter $\eta_0$ is a measure of the recoil experienced by an atom after emission (or absorption) of a photon of wavelength $\lambda_0$. Finally, the ratio $\eta_0/\xi_{ij}=\ell_0/r_{ij}$ is a quantifier of the overlap between atomic wave packets (see Fig.~\ref{fig:gaussians}). Equation~(\ref{gijgsdis}) is remarkable in that it is valid for any values of both $\xi_{ij}$ and $\eta_0$. It provides an accurate description of the combined effects of indeterminacy in atomic positions and recoil on the dissipitave dynamics of atoms for any possible realizations of the three characteristic lengths $r_{ij}$, $\ell_0$ and $\lambda_0$. In particular, it allows for a full description of recoil effects beyond the Lamb-Dicke regime, i.e.\ when $\eta_0 \gtrsim 1$. Table~\ref{table} summarizes several regimes that our theory covers as defined through the comparison of the adimensional parameters $\xi_{ij}$ and $\eta_0$.

\begin{table}
\renewcommand{\arraystretch}{1.6}
\begin{center}
\begin{tabular}{|c|c|}
\hline
Regime & Relevant Phenomena \\
\hline
\hline
 \, $1 \lesssim \eta_0 \ll \xi_{ij}$\, & recoil effects\\
\hline
\,$\eta_0 \lesssim \xi_{ij} \ll 1$\, & cooperative effects \\
\hline
 \,$1 \ll \xi_{ij} \lesssim \eta_0$ \, & indistinguishability, recoil effects\\
\hline
 \multirow{2}*{ \,$\xi_{ij} \ll 1  \lesssim \eta_0$\, } & cooperative effects, recoil effects,\\[-4pt]
 & indistinguishability\ \,  \\
\hline
 \,$\xi_{ij} \lesssim \eta_0 \ll 1$\,   & \, cooperative effects, indistinguishability\, \\
\hline
\end{tabular}
\end{center}
\caption{Regimes and relevant phenomena related to the different ranges of the adimensional parameters $\xi_{ij}=k_0 r_{ij}$ and $\eta_0=k_0 \ell_0$. Recoil effects result from photon emission processes and are significant when $\eta_0 \gg 1$. Cooperative effects reflect the fact that the atoms do not behave as independent emitters when $\xi_{ij} \ll 1$ (provided $\eta_0$ is not too large, see Fig.~\ref{fig:ggm}). Indistinguishability becomes significant as soon as the wave packets overlap (i.e.\ when $\xi_{ij} \lesssim \eta_0$).
}\label{table}
\end{table}

Let us consider two important limiting cases : I. when the distance between any two atoms is much larger than $\lambda_0$ ($\xi_{ij} \gg 1$ : no cooperative effects in the case of classical positions), and II. when
the distance between any two atoms is much smaller than $\lambda_0$ ($\xi_{ij} \ll 1$ : superradiant regime). In the regime I, Eq.~(\ref{gijgsdis}) reduces to
\begin{equation}
\gamma_{ij}^\mathrm{sep} \stackrel[\xi_{ij} \gg 1]{}{\simeq} \,\frac{3\gamma_0 }{2}\, p_{ij} \,\frac{\sin\xi_{ij}}{\xi_{ij}}\,e^{-\eta_0^2}\label{gijfar}
\end{equation}
with $p_{ij}$ the angular factor given by Eq.~(\ref{pqpi}) for a $\pi$ transition and by Eq.~(\ref{pqsigma}) for a $\sigma^\pm$ transition.
This result differs by a factor $e^{-\eta_0^2}$ from the classical result that is obtained for atoms at fixed positions (i.e.\ the radiative term of Eq.~(\ref{gammaijcl})). This factor arises from the quantization of the atomic motion and can be interpreted as a reduction of phase coherence in the cooperative emission due to the uncertainty in the atomic positions.  It is reminiscent of the Debye-Waller factor $\exp\left(-k_B T k_0^2/3M\Omega^2\right)$ typical for neutron scattering, where the position of the atoms is smeared out due to their thermal motion~\cite{Deb13} (here $T$ is the temperature, $k_B$ the Boltzmann constant, $M$ the atomic mass, $\Omega$ the atomic oscillation frequency and $k_0$ the neutron wavenumber).
In the opposite regime ($\xi_{ij}\ll 1$) and for any Lamb-Dicke parameter $\eta_0$, Eq.~(\ref{gijgsdis}) reduces to
\begin{multline}
\label{gijgscsran}
\gamma^{\mathrm{sep}}_{ij} \stackrel[\xi_{ij} \ll 1]{}{\simeq} \gamma_0  \left[ \sqrt{\pi}\, \mathrm{erf}\left(\eta_0\right) \frac{ (8-2q_{ij}) \eta_0^2 + 3 q_{ij} }{16 \eta_0^3}\right.\\
\left.  - \frac{3q_{ij}\, e^{-\eta_0^2} }{8 \eta_0^2}\right].
\end{multline}
In particular, in the Lamb-Dicke regime ($\eta_0\ll 1$), the decay rates decrease with $\eta_0$ as
\begin{equation}
\gamma_{ij}^{\mathrm{sep}}\stackrel[\eta_0 \ll 1]{}{\simeq} \gamma_0 \left( 1 - \frac{5+q_{ij}}{15} \eta_0^2 \right),
\end{equation}
while for large values of $\eta_0$, we have
\begin{equation}
\gamma_{ij}^{\mathrm{sep}}\stackrel[\eta_0 \gg 1]{}{\simeq} \gamma_0 \,\frac{\sqrt{\pi}(4-q_{ij})}{8\eta_0}.
\end{equation}

\begin{figure}
\begin{center}
\includegraphics[width=0.45\textwidth]{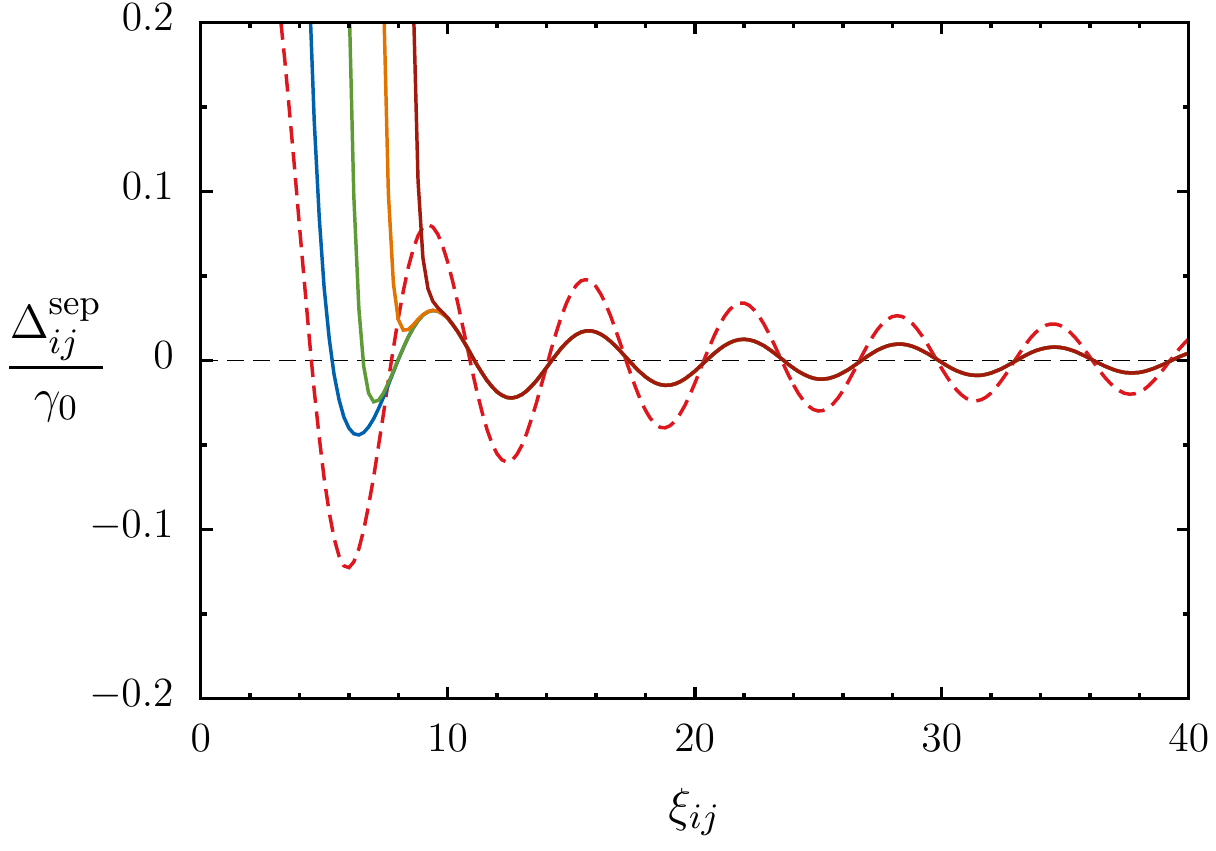}
\caption{(Color online) Regularized dipole-dipole shifts $\Delta_{ij}^\mathrm{sep}$ as a function of $\xi_{ij}=k_0r_{ij}$ for atoms in Gaussian states (\ref{Gaussianstate}) with a Lamb-Dicke parameter $\eta_0 = 1$ and different cutoffs (solid lines from left to right) : $k_0\epsilon = 10^{-1}$ (blue curve), $k_0\epsilon = 10^{-2}$ (green curve), $k_0\epsilon = 10^{-3}$ (orange curve), $k_0\epsilon = 10^{-4}$ (dark red curve). The red dashed curve corresponds to the classical dipole-dipole shift $\Delta^\mathrm{cl}$ given by Eq.~(\ref{deltaijcl}). The plots shown are for a $\pi$ transition with $\alpha_{ij}=\pi/2$.}
\label{fig:dij_gs_eps}
\end{center}
\end{figure}

We now turn to the calculations of the dipole-dipole shifts. Equation (\ref{FinvCijex}) yields for the inverse Fourier transform of (\ref{corkmGSdis})
\begin{equation}\label{invFouTra}
\mathcal{F}^{-1}_{\mathbf{r}'} \left[\mathcal{C}_{ij}^\mathrm{ex,sep} \left(\mathbf{k}'\right)\right] = \frac{e^{- (z' + z'_{ij})^2/4\ell^2_0}}{2\sqrt{\pi}\,\ell_0} \delta(x') \delta(y')\,
\end{equation}
so that Eq.~(\ref{deltaijconv2}) yields
\begin{equation}\label{deltaij_gs}
\Delta_{ij}^\mathrm{sep}(\mathbf{r}_{ij},\ell_0) = \int_{-\infty}^{+\infty} e^{- (z' + z'_{ij})^2/4\ell_0^2} \: \Delta^\mathrm{cl}\left(0,0, z'\right) \,\frac{dz'}{2\sqrt{\pi}\,\ell_0}
\end{equation}
with $\Delta^\mathrm{cl}$ given by Eq.~(\ref{deltaijcl}). Equation~(\ref{deltaij_gs}) depends parametrically on the vector $\mathbf{r}'_{ij}=(0,0,z'_{ij})$  connecting the center of the two Gaussian wave packets. The integral diverges unless a cutoff $\epsilon$ is introduced in order to remove the small values of $z'$ around $z'=0$. Therefore, we introduce the regularized dipole-dipole shifts
\begin{multline}\label{deltaijdisGS}
\Delta_{ij}^\mathrm{sep}(\mathbf{r}_{ij},\ell_0,\epsilon) = \left[\int_{-\infty}^{-\epsilon} + \int_{\epsilon}^{+\infty}\right] e^{- (z' + 
z'_{ij})^2/4\ell_0^2} \, \\
\times \Delta^\mathrm{cl}\left(0,0, z'\right) \, \frac{dz'}{2\sqrt{\pi}\,\ell_0}.
\end{multline}
In Fig.~\ref{fig:dij_gs_eps}, we show the result of a numerical integration of (\ref{deltaijdisGS}) as a function of $\xi_{ij}$ for different cutoffs $\epsilon$. For $\xi_{ij}\gtrsim 10$, all curves are seen to collapse to a single curve displaying similar oscillations as the classical dipole-dipole shift but with a reduced amplitude (depending on the Lamb-Dicke parameter). The chosen cutoffs have no influence in this parameter range. For $\xi_{ij}\lesssim 10$, the curves corresponding to different cutoffs start to differ. The ones with smaller values of $\epsilon$ diverge more rapidly as $\xi_{ij}$ decreases. However, the cutoff cannot be arbitrary small since atoms are not point-like particles but have a finite spatial extent of the order of the Bohr radius $a_0$. For frequencies in the optical domain, this leads to the condition $k_0 \epsilon  > k_0\, a_0 \sim 10^{-3}$. 

\subsubsection{Indistinguishable atoms}

The motional correlation function for indistinguishable atoms in single-atom Gaussian states [see Eq.~(\ref{Gaussianstate})] is obtained by inserting (\ref{overlapGS}) into (\ref{cijexpm}). 
The decay rates (\ref{gijgen}) and dipole-dipole shifts (\ref{deltaijconv2}) for indistinguishable atoms are thus given by
\begin{align}
\label{gijGGpm}
\gamma_{ij}^{\pm}(\{\mathbf{r}_{ij}\},\ell_0) &= \sum_{\pi,\pi'}w^{\pi\pi',\pm} \:\gamma^\mathrm{sep}_{ij}\big(\bar{\mathbf{r}}_{ij}^{\pi\pi'},\ell_0\big),\\
\label{deltaijpmGS}
\Delta_{ij}^{\mathrm{\pm}}(\{\mathbf{r}_{ij}\},\ell_0)  &=  \sum_{\pi, \pi'} \, w^{\pi\pi',\pm}\,\Delta_{ij}^\mathrm{sep}\big(\bar{\mathbf{r}}_{ij}^{\pi\pi'},\ell_0\big),
\end{align}
with
\begin{equation}\label{npmGSz}
w^{\pi\pi',\pm} = \frac{\displaystyle s_{\pm}^{\pi}\,s_{\pm}^{\pi'} \prod_{n = 1}^N e^{-\frac{z'^2_{\pi(n)\pi'(n)}}{8\ell_0^2}}}{\displaystyle
\sum_{\tilde{\pi},\tilde{\pi}'}s_{\pm}^{\tilde{\pi}}\,s_{\pm}^{\tilde{\pi}'} \prod_{n = 1}^N e^{-\frac{z'^2_{\tilde{\pi}(n)\tilde{\pi}'(n)}}{8\ell_0^2}}},
\end{equation} 
where $\gamma^\mathrm{sep}_{ij}$ and $\Delta_{ij}^\mathrm{sep}$ are given by Eqs.~(\ref{gijgsdis}) and (\ref{deltaijdisGS}) respectively, and
\begin{equation}
\bar{\mathbf{r}}_{ij}^{\pi\pi'}=\frac{1}{2}\left(\mathbf{r}_{\pi(i)\pi(j)} + \mathbf{r}_{\pi'(i)\pi'(j)}\right).
\end{equation} 
Equations (\ref{gijGGpm}) and (\ref{deltaijpmGS}) now depend on all $\mathbf{r}_{ij}$, but are equal for all $i$ and $j$ as a consequence of indistinguishability, as discussed in the previous section.

\begin{figure*}
\begin{center}
\includegraphics[width=\textwidth]{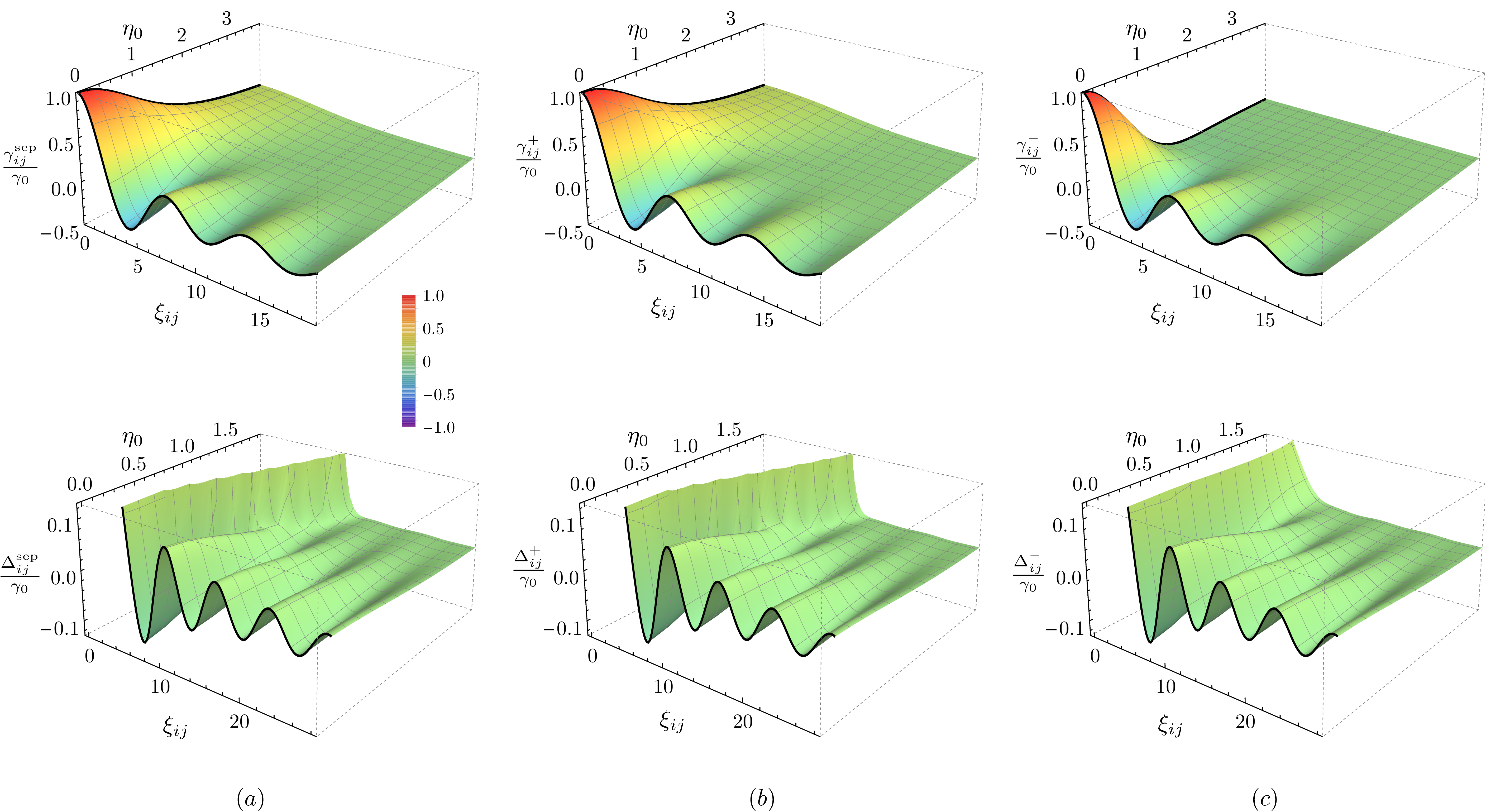}
\end{center}
\caption{(Color online) Off-diagonal decay rates (top) and regularized dipole-dipole shifts (bottom) for the configuration illustrated in Fig.~\ref{fig:gaussians} as a function of $\xi_{ij}=k_0 r_{ij}$ and $\eta_0=k_0\ell_0$ for $(a)$ distinguishable atoms [Eq.~(\ref{gijgsdis}) and (\ref{deltaijdisGS})] and $(b), (c)$ indistinguishable atoms [Eq.~(\ref{gijGGpm}) and (\ref{deltaijpmGS}) for $N=2$]. The plots shown are for a $\pi$ transition with $\alpha_{ij}=\pi/2$ and a cutoff $k_0\epsilon=0.01$. The black solid curves at the front of each plot (for $\eta_0=0$) are the classical decay rate (\ref{gammaijcl}) and dipole-dipole shift (\ref{deltaijcl}). The black solid curves on the left of each plot of the decay rates (for $\xi_{ij}=0$) correspond, in the cases $(a)$ and $(b)$, to Eq.~(\ref{gijgscsran}).} \label{fig:ggm}
\end{figure*}

Figure~\ref{fig:ggm} displays the decay rates and regularized dipole-dipole shifts as a function of $\xi_{ij}$ and $\eta_0$, both for $(a)$ distinguishable and $(b),(c)$ indistinguishable atoms (corresponding to symmetric and antisymmetric wave functions respectively). The decay rates $\gamma_{ij}^\mathrm{sep}$ and $\gamma_{ij}^\pm$ are those given in Eqs.~(\ref{gijgsdis}) and (\ref{gijGGpm}), while the dipole-dipole shifts $\Delta_{ij}^\mathrm{sep}$ and $\Delta_{ij}^\pm$ are those given in Eqs.~(\ref{deltaijdisGS}) and (\ref{deltaijpmGS}). In the Lamb-Dicke regime ($\eta_0 \ll 1$), the quantum fluctuations of the atomic positions are small and the decay rates and dipole-dipole shifts only slightly depart from their classical values, Eqs.~(\ref{gammaijcl}) and (\ref{deltaijcl}). Beyond the Lamb-Dicke regime ($\eta_0 \gtrsim 1$), the decay rates and dipole-dipole shifts still display oscillations as a function of $\xi_{ij}$ but with a reduced amplitude. This reduction in amplitude is more and more pronounced as $\eta_0$ increases. Physically, this can be understood as the result of an average over the atomic positions at the scale of the atomic wave packets of the corresponding oscillating classical quantities (see Eqs.~(\ref{gammaijconv}) and (\ref{deltaijconv2})). For small interatomic distances in comparison to the wave packets extension ($\xi_{ij} \lesssim \eta_0$), the symmetry of the wave function has major effects on how fast the amplitude decreases with $\eta_0$. It is seen to decrease much faster for the antisymmetric wave function than for the symmetric one (see $(b)$ and $(c)$ in the middle panel). Symmetric and separable wave functions yield very close results because their two-atom reduced density matrices $\rho_{ij}^+$ and $\rho_{ij}^{\mathrm{sep}}$ are very close. In particular, when $\xi_{ij} \to 0$, we have $\rho_{ij}^+ \to \rho_{ij}^\mathrm{sep}$ and $\gamma_{ij}^+\to\gamma_{ij}^\mathrm{sep}$ with $\gamma_{ij}^\mathrm{sep}$ given by Eq.~(\ref{gijgscsran}). On the contrary, the two-atom reduced density matrix $\rho_{ij}^-$ differ significantly because of the Pauli exclusion principle. When the overlap between atomic wave packets becomes negligible ($\xi_{ij} \gg \eta_0$), the decay rates and the dipole-dipole shifts are approximately equal for the symmetric, antisymmetric and separable wave functions, showing that atoms can be treated as distinguishable particles in this regime. 

\subsection{Harmonic oscillator eigenstates}
We now consider as single-atom motional states the vibrational states of harmonically trapped atoms centered around the positions $\mathbf{r}'_j$ ($j = 1, \dotsc, N$), hereafter referred as Fock states. We denote them by $\ket{\phi_{(n,\mathbf{r}'_j)}}$ where $n = 0, 1, \dotsc$ stands for the number of vibrational excitations. Gaussian states are a particular case ($n = 0$) of this more general class of states. As previously, atoms are taken to be aligned along the $z'$-direction and their motion is quantized only along this direction.

In the position representation, the single-atom motional Fock states $|\phi_{(n,\mathbf{r}'_j)} \rangle $ with typical size $\ell_0$ along $z'$ are given by
\begin{equation}\label{fockstater}
\phi_{(n,\mathbf{r}'_j)}(\mathbf{r}') = \frac{e^{-\left(z' - z'_j\right)^2/4\ell_0^2}}{\left(2^n n!\right)^\frac{1}{2}\left(2\pi \ell^2_0\right)^\frac{1}{4}}\,H_n\left(\frac{z'-z'_j}{\sqrt{2}\ell_0}\right) \delta\left(x'_j\right)\delta\left(y'_j\right)
\end{equation}
where $H_n(z')$ is the Hermite polynomial of order $n$. The overlap integral (\ref{overlap}) between two Fock states at the same position $\mathbf{r}'$ reads~\cite{Win79}
\begin{multline}\label{overlapFS}
I_{(n_i,\mathbf{r}') (n_j,\mathbf{r}')}(\mathbf{k}') = e^{i \mathbf{k}' \boldsymbol{\cdot} \mathbf{r}'}\, e^{-k'^2_{z'} \ell_0^2/2} \\
 \times \sqrt{\frac{n_{<}!}{(n_{<}+\Delta n)!}}\, \big(i k'_{z'} \ell_0\big)^{\Delta n}\,L^{\Delta n}_{n_{<}}\big(k_{z'}'^2 \ell_0^2\big)
\end{multline}
where $L^{\alpha}_{n}$ are the generalized Laguerre polynomials of degree $n$, $\Delta n=|n_i-n_j|$ and $n_{<}=\mathrm{min}\{n_i,n_j\}$.

\subsubsection{Distinguishable atoms}

\begin{figure}
\begin{center}
\includegraphics[width=0.47\textwidth]{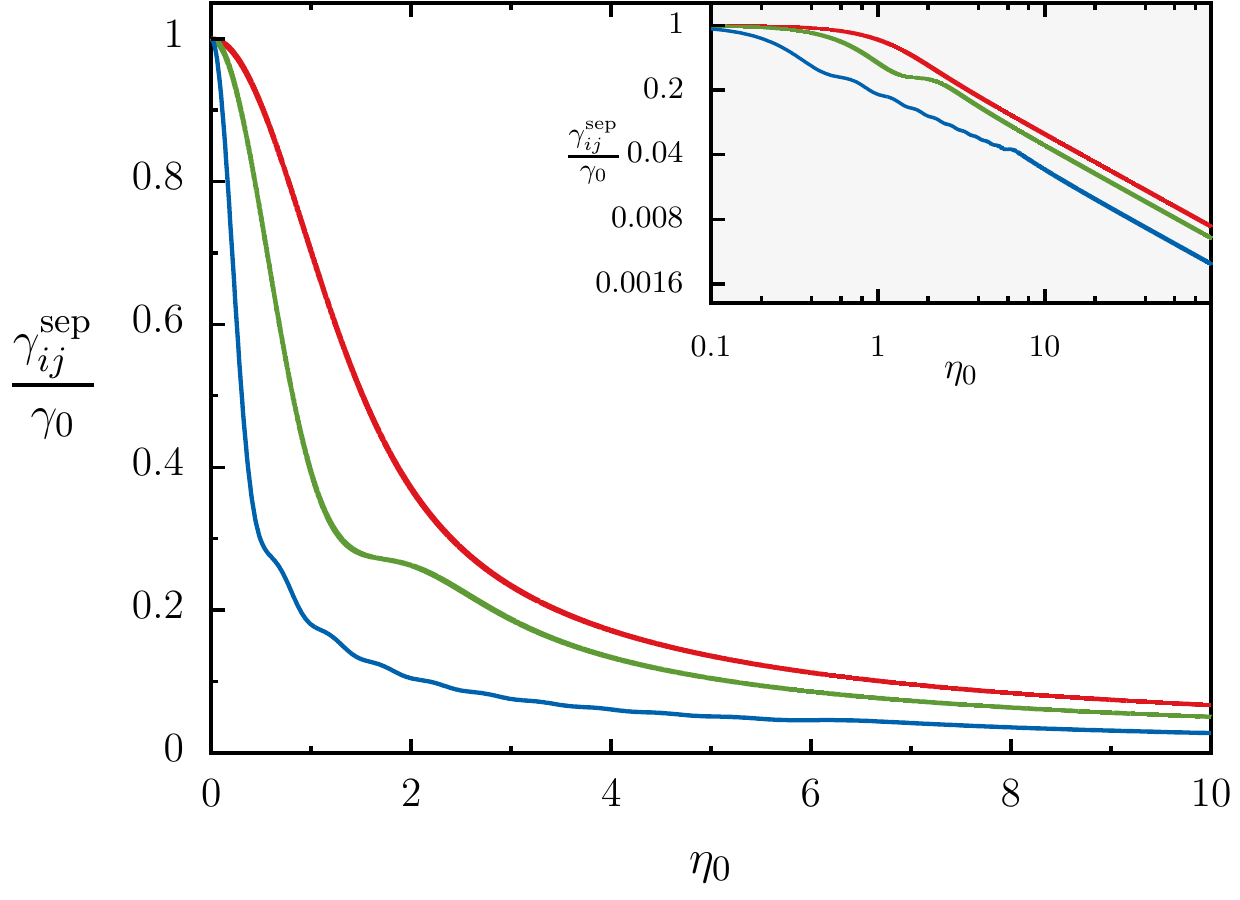}
\end{center}
\caption{(Color online) Decay rates in the superradiant regime ($\xi_{ij}\ll 1$) as a function of the Lamb-Dicke parameter $\eta_0$ for atoms $i$ and $j$ initially in the same motional Fock state $|\phi_{(n, \mathbf{0})}\rangle$ centered around the origin with number of vibrational excitations $n=0$ (ground state - red curve), $n=1$ (green curve) and $n=10$ (blue curve). In this situation, $\gamma_{ij}^{\mathrm{sep}}$ and $\gamma_{ij}^{+}$ coincide. Inset: same figure in log-log scale showing the power-law decrease of $\gamma_{ij}^{\mathrm{sep}}$ as $1/\eta_0$. The plots shown are for a $\pi$ transition with $\alpha_{ij}=\pi/2$.} \label{gammaij_fock}
\end{figure}

When atom $i$ is in the state $\ket{\phi_{(n_i,\mathbf{r}'_i)}}$ and atom $j$ in the state $\ket{\phi_{(n_j,\mathbf{r}'_j)}}$, according to Eq.~(\ref{overlapFS}), the correlation function (\ref{cijexdis}) reads 
\begin{align}
\mathcal{C}_{ij}^{\mathrm{ex,sep}}(\mathbf{k}') &= I_{(n_i,\mathbf{r}'_i)(n_i,\mathbf{r}'_i)}(\mathbf{k}')\,I_{(n_j,\mathbf{r}'_j)(n_j,\mathbf{r}'_j)}(-\mathbf{k}') \nonumber\\[5pt]
&= e^{i \mathbf{k}'\boldsymbol{\cdot} \mathbf{r}'_{ij}}\, e^{- k'^2_{z'} \ell_0^2} \,L^{0}_{n_i}\big(k_{z'}'^2 \ell_0^2\big) \,L^{0}_{n_j}\big(k_{z'}'^2 \ell_0^2\big). \label{corkmFSdis}
\end{align}
The decay rates of distinguishable atoms with Fock states at arbitrary positions can be obtained by inserting Eq.~(\ref{corkmFSdis}) into Eq.~(\ref{gijgen}) and performing the angular integration. Simple analytical expressions can be obtained in the limit $\xi_{ij}\to 0$ (superradiant regime). To this end, we first express the product of Laguerre polynomials as a linear combination of these same polynomials, 
\begin{equation}\label{laguerreproduct}
L_{n_i}^0(x) L_{n_j}^0(x)=\sum\limits_{\ell=|n_i-n_j|}^{n_i+n_j}c_{n_i,n_j,\ell} \,L_{\ell}^0(x)
\end{equation}
with
\begin{equation}\label{laguerreproductcoeff}
c_{n_i,n_j,\ell}=\left(-\frac{1}{2}\right)^p\sum_n\frac{2^{2n} (n_i+n_j-n)!}{(n_i-n)!(n_j-n)!(2n-p)!(p-n)!},
\end{equation}
where $p=n_i+n_j-\ell$ and the sum over $n$ runs over all integers such that the arguments of the factorials are positive \cite{Gil60}. By plugging $\mathcal{C}_{ij}^{\mathrm{ex,sep}}(\mathbf{k}_0')$ with $e^{i \mathbf{k}_0' \boldsymbol{\cdot}\mathbf{r}'_{ij}} \approx 1$ into Eq.~(\ref{gijgen}) and performing the integration over all directions, we get
\begin{widetext}
\begin{equation}\label{gij_fs_sol}
\gamma_{ij}^\mathrm{sep}(n_i, n_j,\ell_0) \stackrel[\xi_{ij} \ll 1]{}{\simeq} \frac{\gamma_0}{4} \sum\limits_{\ell=|n_i-n_j|}^{n_i+n_j}c_{n_i,n_j,\ell} 
\Bigg[  q_{ij} \; {}_2F_2\left(\frac{3}{2},\ell+1;1,\frac{5}{2};-\eta_0^2 \right) \Bigg. 
 \Bigg. +  (4 - q_{ij}) \; {}_2F_2\left(\frac{1}{2},\ell+1;1,\frac{3}{2};-\eta_0^2\right) \Bigg]
\end{equation}
\end{widetext}
with $q_{ij}$ given by Eq.~(\ref{pqpi}) for $\pi$ transition and by Eq.~(\ref{pqsigma}) for $\sigma^\pm$ transition and ${}_pF_q(\mathbf{a};\mathbf{b};z)$ the generalized hypergeometric series~\cite{Abr70,footnote7}.
The decay rates (\ref{gij_fs_sol}) for atoms in the same Fock state are shown in Fig.~\ref{gammaij_fock} as a function of the Lamb-Dicke parameter for a $\pi$ transition with $\alpha_{ij} = \pi/2$ and for different excitation numbers $n=n_i=n_j$. At fixed Lamb-Dicke parameter, the decay rates are smaller as the excitation number increases. For large Lamb-Dicke parameters, they decrease like a power-law, as can be seen from the inset. Some oscillations are present for excitation numbers $n>0$, which we attribute to oscillations (in momentum space) of the motional wave packets.

\subsubsection{Indistinguishable atoms}

For indistinguishable atoms in Fock states, the correlation function is given by Eq.~(\ref{cijexpm}). The decay rates can be evaluated for arbitrary positions and Lamb-Dicke parameter by inserting (\ref{cijexpm}) into (\ref{gijgen}). In the limit $\xi_{ij} \to 0$, the vibrational states for different excitation numbers are orthogonal and the decay rates are given by
\begin{multline}
\label{FockN}
\gamma_{ij}^{\pm}(\{n_i\},\ell_0) \stackrel[\xi_{ij} \ll 1]{}{\simeq} \frac{1}{N!} \sum_{\pi,\pi'} s_{\pm}^{\pi}\,s_{\pm}^{\pi'} \,\sigma_{ij}^{\pi,\pi'} \\
\times \int \sum_{\eps} \gijemkz \,  I_{\pi(i)\pi'(i)}(\mathbf{k}_0) \, I_{\pi(j)\pi'(j)}(- \mathbf{k}_0)  \,\frac{d\Omega}{(2\pi)^2},
\end{multline}
with $\gijemkz$ and $I_{\alpha \beta}(\mathbf{k}_0)$ given by Eqs.~(\ref{gijclass}) and (\ref{overlapFS}) and
\begin{equation}
\sigma_{ij}^{\pi,\pi'} =\prod_{n = 1 \atop n \neq i,j}^N \!\delta_{\pi(n)\pi'(n)},
\end{equation}
where $\delta_{\pi(n)\pi'(n)}$ is the Kronecker symbol. For equal excitation numbers, the symmetric motional state $\rho_{ij}^+$ becomes separable in this regime and the symmetric decay rates $\gamma_{ij}^+$ tend to $\gamma_{ij}^\mathrm{sep}$ given by Eq.~(\ref{gij_fs_sol}).

\subsection{Thermal states in a harmonic trap}

We now consider as motional state the thermal state of atoms trapped in a harmonic potential of frequency $\Omega_{z'}= \hbar/2M\ell_0^2$ along the $z'$ direction. In this case, all atoms occupy the same motional mixed state~\cite{footnote8}
\begin{equation}\label{rhotherm}
\rho_{(\bar{n},\mathbf{0}')}=\sum_{n=0}^{+\infty}\frac{\bar{n}^n}{(1+\bar{n})^{n+1}}\,|\phi_{(n,\mathbf{0}')}\rangle\langle \phi_{(n,\mathbf{0}')}|
\end{equation}
where $\bar{n}=1/\big(e^{\hbar\Omega_{z'}/k_B T}-1\big)$ is the mean phonon number at temperature $T$.
The overlap
\begin{equation}
\label{overlapTS}
 I_{(\bar{n},\mathbf{0}')(\bar{n},\mathbf{0}')}(\mathbf{k}') =\big\langle e^{i k'_{z'} \hat{z}'_j} \big \rangle
= \tr \big( e^{i k'_{z'} \hat{z}'_j} \rho_{(\bar{n},\mathbf{0}')}\big)
\end{equation}
can be evaluated analytically by writing the position operator as $\hat{z}'_j = \ell_0(b_j + b_j^\dagger)$ with $b_j$ and $b_j^\dagger$ the annihilation and creation operators of a motional excitation for atom $j$. Upon using the identity $\big\langle \exp\big({\ell_0 (b_j^\dagger + b_j)}\big)\big\rangle =  \exp\big({\ell_0^2\langle (b_j^\dagger + b_j)^2 \rangle}\big)$ where the expectation value is taken in a thermal state~\cite{Hua01}, we get
\begin{equation}
\label{overlapTS2}
 I_{(\bar{n},\mathbf{0}')(\bar{n},\mathbf{0}')}(\mathbf{k}') =  e^{-k_{z'}'^2 \ell_0^2\left(2\bar{n} + 1\right)/2}.
\end{equation}
The corresponding correlation function (\ref{cijexdis}) reads 
\begin{align}
\mathcal{C}_{ij}^\mathrm{ex, sep}(\mathbf{k}')
= e^{-k_{z'}'^2 \ell_0^2 (2\bar{n}+1)}, \label{corkm}
\end{align}
and is of the same form as for Gaussian states centered around the origin (see Eq.~(\ref{corkmGSdis})), now with a width $\tilde{\ell}_0 = \ell_0 \sqrt{2\bar{n}  +1}$ which depends on the temperature through $\bar{n}$. As a consequence, the decay rates and the dipole-dipole shifts for atoms in the same thermal state are given by Eqs.~(\ref{gijgsdis}) and (\ref{deltaijdisGS}) with $\eta_0$ replaced by $\tilde{\eta}_0 =  \eta_0 \sqrt{2\bar{n} +1}$. The increase in Lamb-Dicke parameter from $\eta_0$ to $\tilde{\eta}_0$ comes from the Debye-Waller factor $e^{-k_0^2\langle \hat{z}'^2_j \rangle }$ where $\langle \hat{z}'^2_j  \rangle =\ell_0^2 (2 \bar{n} + 1)$ is the mean square displacement of atom $j$.


\section{Conclusions}

In this work, we derived a general master equation for the internal dynamics of atoms coupled to the electromagnetic field in vacuum, taking into account the quantization of their motion. Our master equation provides an accurate description of recoil effects, even beyond the Lamb-Dicke regime, and applies equally well to distinguishable and indistinguishable atoms. We obtained general expressions for the dipole-dipole shifts and the decay rates, which determine the conservative and dissipative atomic internal dynamics, in terms of their classical expressions and the motional correlation function defined for arbitrary motional states. We showed that the motional state allows one to engineer the dipole-dipole shifts and the decay rates, and can lead to a large modification compared to the classical value. In particular, we obtained analytical expressions for the decay rates for Gaussian states, harmonic oscillator eigenstates and thermal states, that are relevant in cold atom experiments. 

\begin{acknowledgments}
FD would like to thank the F.R.S.-FNRS for financial support.
FD is a FRIA grant holder of the Fonds de la Recherche Scientifique-FNRS.
\end{acknowledgments}


\begin{thebibliography}{100}
\bibitem{footnote1} We emphasize the quantum treatment of the atomic motion by writing the center-of-mass position and momentum operators $\hat{\mathbf{r}}$ and $\hat{\mathbf{p}}$, in contrast to the classical center-of-mass position and momentum which are denoted by $\mathbf{r}$ and $\mathbf{p}$ respectively.
\bibitem{footnote1b} This choice is perfectly justified in the optical domain since optical modes of the electromagnetic field in thermal equilibrium at a temperature below 300 K are essentially in the ground state.
\bibitem{footnote2} If this is not the case,  it is always possible to add
the term $\sum_{j,\omega}\langle B_{j}\rangle_B A_{j}^{\mathrm{in}}(\omega)\otimes
\mathbb{1}_B$ to $H_A^{\mathrm{in}}$ - which displace the energy levels of the system -
and substract it from $H_{AF}$ such that it holds.
\bibitem{footnote3} In this work, we use for the Fourier transform $\mathcal{F}_\mathbf{k}\left[ \cdot\right]$ and its inverse $\mathcal{F}_\mathbf{r}^{-1}[\cdot]$ the convention
\begin{equation*}
\begin{aligned}
&\ftq{f} = \int_{\mathbb{R}^3}  e^{-i \mathbf{k} \boldsymbol{\cdot} \mathbf{r}} \, f(\mathbf{r})\, d\mathbf{r}, \\
&\ftxiij{g} = \int_{\mathbb{R}^3}  e^{i  \mathbf{k}\boldsymbol{\cdot} \mathbf{r}} \,g(\mathbf{k})\, \frac{d\mathbf{k}}{(2\pi)^3}. \\
\end{aligned}
\end{equation*}
\bibitem{footnote4} The principal value arising here is a mere consequence of second order perturbation theory since resonant intermediate transitions must be discarded~\cite{Don15}.
\bibitem{footnote5} Note that this is a consequence of the second-order perturbative treatment for the atom-field interaction. As Van der Waals dipole-dipole interactions between two atoms in the same internal state result from a fourth-order perturbative development, they are not considered in this work.
\bibitem{footnote7} 
The generalized hypergeometric series is defined by
\begin{equation*}
{}_pF_q(\mathbf{a};\mathbf{b};z)
= \sum_{k=0}^{+\infty} \frac{(a_1)_k\ldots (a_p)_k}{(b_1)_k\ldots (b_q)_k} \, \frac {z^k} {k!}
\end{equation*}
with $(\boldsymbol{\cdot})_k$ the Pochhammer symbol defined as $(a)_0 = 1$ and 
\begin{equation*}
(a)_k =a(a+1)(a+2)\cdots(a+k-1)=\frac{(a+k-1)!}{(a-1)!}.
\end{equation*}
\bibitem{footnote8} There is no contradiction with the fact that we consider the e.m.\ field in vacuum. Indeed, since the motional frequency $\Omega_{z'}$ is much smaller than the field frequencies $\omega \sim \omega_0$ which contribute significantly to the dynamics, the motional and the e.m.\ field states can be taken respectively thermaly excited and in vacuum at the same time.
\end{thebibliography}
\end{document}